\documentclass[twocolumn]{aastex631}

\usepackage{natbib}
\usepackage{epsfig}
\usepackage{graphicx}
\usepackage{subfigure}
\usepackage{float}
\usepackage{amsmath}
\usepackage{color}
\usepackage{amssymb}
\usepackage{apjfonts}

\newcommand{\PMO}{Purple Mountain Observatory, Chinese Academy of Sciences, Nanjing 210023, China}
\newcommand{\USTC}{School of Astronomy and Space Sciences, University of Science and Technology of China, Hefei 230026, China}
\newcommand{\UA}{Department of Physics, The Applied Math Program, and Department of Astronomy,
The University of Arizona, AZ 85721, USA}

\shortauthors{Wei \& Melia}

\graphicspath{{./}{figures/}}

\usepackage{amsmath}
\begin{document}

\title{Investigating Cosmological Models and the Hubble Tension using Localized Fast Radio Bursts}


\correspondingauthor{Jun-Jie Wei}
\email{jjwei@pmo.ac.cn, fmelia@email.arizona.edu}

\author{Jun-Jie Wei}
\affiliation{\PMO}
\affiliation{\USTC}

\author{Fulvio Melia \thanks{John Woodruff Simpson Fellow.}}
\affiliation{\UA}

\begin{abstract}
We use the dispersion measure (DM) and redshift measurements of 24 localized fast radio
bursts (FRBs) to compare cosmological models and investigate the Hubble tension.
Setting a flat prior on the DM contribution from the Milky Way's halo,
$\mathrm{DM_{halo}^{MW}}\in[5,\;80]\;\mathrm{pc\;cm^{-3}}$, the best fit for flat
$\Lambda$CDM is obtained with a Hubble constant
$H_0=95.8^{+7.8}_{-9.2}\;\mathrm{km\;s^{-1}\;Mpc^{-1}}$ and a median matter density
$\Omega_{\mathrm{m}}\approx0.66$. The best fit for the $R_{\mathrm{h}}=ct$ universe
is realized with $H_0=94.2^{+5.6}_{-6.2}\;\mathrm{km\;s^{-1}\;Mpc^{-1}}$. We emphasize
that the $H_0$ measurement depends sensitively on the $\mathrm{DM_{halo}^{MW}}$ prior.
Since flat $\Lambda$CDM has one more free parameter, $R_{\mathrm{h}}=ct$ is
favored by the Bayesian Information Criterion (BIC) with a likelihood of $\sim73\%$
versus $\sim27\%$. Through simulations, we find that if the real cosmology is $\Lambda$CDM,
a sample of $\sim1,150$ FRBs in the redshift range $0<z<3$ would be sufficient to rule out
$R_{\mathrm{h}}=ct$ at a $3\sigma$ confidence level, while $\sim550$ FRBs would be necessary
to rule out $\Lambda$CDM if the real cosmology is instead $R_{\mathrm{h}}=ct$.
The required sample sizes are different, reflecting the fact that the BIC imposes
a severe penalty on the model with more free parameters.
We further adopt a straightforward method of deriving an upper limit to
$H_{0}$, without needing to consider the poorly known probability distribution of the
DM contributed by the host galaxy.
The theoretical DM contribution from the intergalactic medium ($\mathrm{DM_{IGM}}$)
at any $z$ is proportional to $H_0$. Thus, requiring the extragalactic
$\mathrm{DM_{ext}}$ to be larger than $\mathrm{DM_{IGM}}$ delimits
$H_0$ to the upside. Assuming flat $\Lambda$CDM, we have
$H_0<89.0\;\mathrm{km\;s^{-1}\;Mpc^{-1}}$ at a 95\% confidence level.
\end{abstract}


\section{INTRODUCTION}
The exact value of the Hubble constant, $H_0$, characterizing the current expansion rate
of the Universe, is one of the most pressing issues in modern cosmology. In the last decade,
the precision of measuring $H_0$ has been greatly improved, but a significant disparity
now appears between its determination at early times and in our local neighbourhood
\citep{2019NatAs...3..891V,2021CQGra..38o3001D}. The near\-by value of $H_0$ measured
from the Cepheid-calibrated type Ia supernovae (SNe Ia) is $73.04\pm1.04$
$\mathrm{km\;s^{-1}\;Mpc^{-1}}$ \citep{2022ApJ...934L...7R}, representing a
$5\sigma$ tension with that inferred from {\it Planck} cosmic microwave background
(CMB) observations in the context of standard $\Lambda$CDM ($H_{0}=67.4\pm0.5$
$\mathrm{km\;s^{-1}\;Mpc^{-1}}$; \citealt{2020A&A...641A...6P}). If this discrepancy
does not arise from unknown systematic uncertainties, new physics beyond $\Lambda$CDM
may be required to alleviate this so-called `Hubble tension'
\citep{Melia2020,2020PhRvD.102b3518V,Melia2022}. It is therefore essential to find other
independent methods of measuring $H_{0}$ to investigate this tension more deeply.

Fast radio bursts (FRBs) are mysterious radio transients with excess dispersion measures (DMs)
with respect to the expected contributions from the Milky Way \citep{Lorimer2007,Thornton2013,
Petroff2019,Platts2019,Xiao2021,2022arXiv221203972Z}. Since the discovery of the first burst
(FRB 20010724) in 2007 \citep{Lorimer2007}, more than 600 FRBs have been detected
\citep{2021ApJS..257...59C}. Of these, 26 have been localized, confirming that
most, if not all, FRBs originate at cosmological distances.

The DM is the integral over the number density of free electrons along
the propagation path between the source and observer, that contains important
information on the cosmological distance. Therefore, the combined DM and redshift
information of FRBs can be used to study cosmology, such as constraining the baryon
number density of the Universe \citep{Deng2014,McQuinn2014,Macquart2020,2022ApJ...940L..29Y},
the dark energy equation of state \citep{Gao2014,Zhou2014,Walters2018,2018ApJ...860L...7W,
2020ApJ...901..130Z,Zhao2020,2022JCAP...02..006Q}, cosmic reionization history
\citep{Deng2014,Zheng2014,Hashimoto2021}, cosmic proper distance \citep{Yu2017}, the baryon
mass fraction in the intergalactic medium (IGM; \citealt{Li2019,Li2020,Walters2019,
Wei2019,2023ApJ...944...50W}), the Hubble constant \citep{Hagstotz2022,2022MNRAS.516.4862J,Wu2022},
and so on. Additionally, strongly lensed FRBs have also been proposed for probing the nature of
compact dark matter \citep{2016PhRvL.117i1301M,2018A&A...614A..50W}, for measuring the Hubble
constant and cosmic curvature \citep{Li2018}, and for determining the post-Newtonian parameter
\citep{2022MNRAS.516.1977G}.

But though FRBs have previously been used to measure $H_0$ in $\Lambda$CDM (see, e.g.,
\citealt{Hagstotz2022,2022MNRAS.516.4862J,Wu2022}), they have not yet been utilized for
model selection---our primary focus in this paper. Here, we assemble the most up-to-date sample
of localized FRBs to examine whether these sources can also be used for comparative studies
between competing models, such as $\Lambda$CDM and an alternative Friedmann-Lema\^itre-Robertson-Walker
cosmology known as the $R_{\rm h}=ct$ universe
\citep{2007MNRAS.382.1917M,2012MNRAS.419.2579M,Melia2020}. In view of the fact that the
DM contribution from the FRB host galaxy, $\mathrm{DM_{host}}$, cannot be determined
exactly from observations, we further adopt a straightforward method of deriving an
upper limit to $H_{0}$ by requiring that the IGM portion of the DM, i.e., $\mathrm{DM_{IGM}}$,
for any FRB at any $z$ must be smaller than the corresponding extragalactic $\mathrm{DM_{ext}}$,
without involving any assumption on the probability distribution for $\mathrm{DM_{host}}$.
Our study provides a consistency test of the low-redshift cosmic expansion,
and possibly suggests improvements required to arbitrate the Hubble tension.

In Section~\ref{sec:method}, we briefly describe
the DM measurements of localized FRBs, and then constrain the cosmological parameters
within the context of $\Lambda$CDM and the $R_{\rm h}=ct$ universe (Section~\ref{sec:model}).
As we shall see, the current FRB sample is not large enough to distinguish
these two models, and we forecast in Section~\ref{sec:MC} how many FRBs will be required from
future detections to rule out one or the other expansion scenario at a $3\sigma$ confidence level.
In Section~\ref{sec:HT}, we introduce the methodology of deriving upper limits on $H_{0}$,
followed by the results of our analysis. We end with our conclusions in Section~\ref{sec:Summary}.

\begin{table*}
	\centering
	\caption{Properties of 26 localized FRBs}
	\label{tab1}
	\begin{tabular}{lcccccr}
		\hline\hline
		Name & Redshift & $\mathrm{DM_{obs}}$  & $\mathrm{DM_{ISM}^{MW}}$ & Repeater & Host Type& Reference$\null\qquad\;$\\
             &          & $(\mathrm{pc\;cm^{-3}})$  & $(\mathrm{pc\;cm^{-3}})$ &    &      \\
		\hline
FRB 20200110E	& $-0.0001$	&$	87.75	\pm	0.05$&	30--40	& Y & II & \cite{Kirsten2022}	\\
FRB 20181030A	&	0.0039	&$	103.5	\pm	0.3	$&	41	& Y & I & \cite{2021ApJ...919L..24B}	\\
FRB 20180916B	&	0.0337	&$	348.76	\pm	0.1	$&	200	& Y & II & \cite{Marcote2020}	\\
FRB 20211127I	&	0.0469	&$	234.83			$&	42.5	& N & III & \cite{2022MNRAS.516.4862J}	\\
FRB 20211212A	&	0.0715	&$	206			$&	27.1	& N & III & \cite{2022MNRAS.516.4862J}	\\
FRB 20201124A	&	0.098	&$	413.52	\pm	0.05	$&	123.2	& Y & II & \cite{2022MNRAS.513..982R}	\\
FRB 20190608B	&	0.1178	&$	338.7	\pm	0.5	$&	37.2	& N & III & \cite{2021ApJ...922..173C}	\\
FRB 20210807D	&	0.12927	&$	251.9			$&	121.2	& N & III & \cite{2022MNRAS.516.4862J}	\\
FRB 20200430A	&	0.16	&$	380.25	\pm	0.4	$&	27	& N & III & \cite{2020ApJ...903..152H}	\\
FRB 20121102A	&	0.19273	&$	557	\pm	2	$&	188	& Y & I & \cite{Chatterjee2017}	\\
FRB 20210117A	&	0.2145	&$	730	$&	34.4	& N & III & \cite{2022MNRAS.516.4862J}	\\
FRB 20191001A	&	0.234	&$	506.92	\pm	0.04	$&	44.7 & N	& III & \cite{2020ApJ...903..152H}	\\
FRB 20190714A	&	0.2365	&$	504.13	\pm	2	$&	38	&  N & III & \cite{2020ApJ...903..152H}	\\
FRB 20190520B	&	0.241	&$	1205	\pm	4	$&	60	& Y & I & \cite{Niu2022}	\\
FRB 20191228A	&	0.2432	&$	297.5	\pm	0.05	$&	33	& N & III & \cite{Bhandari2022}	\\
FRB 20210320C	&	0.2797	&$	384.8			$&	42.2	& N & III & \cite{2022MNRAS.516.4862J}	\\
FRB 20190102C	&	0.291	&$	363.6	\pm	0.3	$&	57.3	& N & III & \cite{2020ApJ...895L..37B}	\\
FRB 20180924B	&	0.3214	&$	361.42	\pm	0.06	$&	40.5 & N	& III & \cite{2019Sci...365..565B}	\\
FRB 20180301A	&	0.3304	&$	536	\pm	8	$&	152	& Y & I & \cite{Bhandari2022}	\\
FRB 20200906A	&	0.3688	&$	577.8	\pm	0.02	$&	36	& N & III & \cite{Bhandari2022}	\\
FRB 20190611B	&	0.378	&$	321.4	\pm	0.2	$&	57.8 & N	& III & \cite{2020ApJ...903..152H}	\\
FRB 20181112A	&	0.4755	&$	589.27	\pm	0.03	$&	102	& N & III & \cite{2019Sci...366..231P} \\
FRB 20190711A	&	0.522	&$	593.1	\pm	0.4	$&	56.4 & Y	& I & \cite{2020ApJ...903..152H}	\\
FRB 20190614D	&	0.6	    &$	959.2	\pm	0.5	$&	83.5	& N & III & \cite{2020ApJ...899..161L}	\\
FRB 20190523A	&	0.66	&$	760.8	\pm	0.6	$&	37	& N & III & \cite{2019Natur.572..352R}	\\
FRB 20220610A	&	1.016	&$	1457.624\pm	0.001$&	31	& N & III & \cite{2022arXiv221004680R}	\\
		\hline
	\end{tabular}
\end{table*}

\section{The properties of localized FRBs}\label{sec:method}
The precise localization of an FRB within its host galaxy allows us to correlate its
$\mathrm{DM_{obs}}$ to a redshift. The $\mathrm{DM_{obs}}$ is a measure of the column
density of free electrons, $n_e$, along the line of sight $l$: $\mathrm{DM_{obs}}=
\int n_{e} \mathrm{d}l/(1+z)$. Physically, we expect $\mathrm{DM_{obs}}$ to have
contributions from four components: two from the Milky Way, i.e., one from the interstellar
medium (ISM) and a second from the halo; and two extragalactic ones, i.e., the IGM and the
FRB host galaxy:
\begin{equation}\label{eq:DMobs}
\begin{split}
  \mathrm{DM_{obs}}(z)&=\mathrm{DM_{MW}}+\mathrm{DM_{ext}}(z)  \\
    &=\mathrm{DM_{ISM}^{MW}}+\mathrm{DM_{halo}^{MW}}+\mathrm{DM_{IGM}}(z)
	+\frac{\mathrm{DM_{host,loc}}}{1+z}\;,
\end{split}
\end{equation}
where the $(1+z)$ factor converts the local DM contribution from the host galaxy, $\mathrm{DM_{host,loc}}$,
to the observed one $\mathrm{DM_{host}}$ \citep{Deng2014}. The DM due to the Milky Way's ISM,
$\mathrm{DM_{ISM}^{MW}}$, can be estimated quite well using Galactic electron density models
\citep{Cordes2002}. The contribution from the Milky Way's halo, $\mathrm{DM_{halo}^{MW}}$,
is not as well constrained, but is expected to range from 50 to 80 $\mathrm{pc\;cm^{-3}}$
\citep{Prochaska2019a}. There are studies suggesting a smaller DM in the Milky
Way halo (e.g.  \citealt{2020MNRAS.496L.106K}), and also an observation of the M81
FRB 20200110E showing $\mathrm{DM_{halo}^{MW}}<32-42$ $\mathrm{pc\;cm^{-3}}$ along that
line of sight \citep{Kirsten2022}. These values are already rejected by the estimate of
\cite{Prochaska2019a}, however. Given the high uncertainty in this term, we shall marginalize
$\mathrm{DM_{halo}^{MW}}$ using a wide flat prior of $[5,\;80]$ $\mathrm{pc\;cm^{-3}}$.
The observed extragalactic dispersion measure, $\mathrm{DM_{ext}}$, can then be derived
by deducting the Milky Way's contributions to $\mathrm{DM_{obs}}$, i.e.,
\begin{equation}\label{eq:DMext}
\begin{split}
  \mathrm{DM_{ext}}&=\mathrm{DM_{obs}}-\mathrm{DM_{ISM}^{MW}}-\mathrm{DM_{halo}^{MW}}\\
                   &=\mathrm{DM_{IGM}}+\mathrm{DM_{host}}\;.
\end{split}
\end{equation}
Moreover, for a well-localized FRB, the $\mathrm{DM_{IGM}}$ value in a certain direction
can in principle be measured by subtracting the host galaxy's portion from $\mathrm{DM_{ext}}$, i.e.,
\begin{equation}
  \mathrm{DM_{IGM}}=\mathrm{DM_{ext}}-\mathrm{DM_{host}}\;.
\end{equation}
The value of $\mathrm{DM_{host}}$ is poorly known, however, as it depends very closely on the type
of host galaxy, the near-source plasma, and the relative orientations of the host and source
\citep{2015RAA....15.1629X}, and may likely range from tens to hundreds of $\mathrm{pc\;cm^{-3}}$.
To model the possible large spread of $\mathrm{DM_{host}}$, we assume that it follows a lognormal
distribution with an asymmetric tail to large values
\citep{Macquart2020}:
\begin{equation}\label{eq:Phost}
	P_{\rm host}\left(\mathrm{DM}_{\mathrm {host}} \right)=\frac{1}{\sqrt{2 \pi}\mathrm{DM}_{\mathrm {host }} \sigma_{\mathrm {host }} } \exp \left[-\frac{\left(\ln \mathrm{DM}_{\mathrm {host}}-\mu_{\mathrm {host}}\right)^2}{2 \sigma_{\mathrm {host}}^{2}}\right]\;,
\end{equation}
where $\mu_{\mathrm{host}}$ and $\sigma_{\mathrm{host}}$ represent, respectively, the mean and standard
deviation of $\ln \mathrm{DM_{host}}$. For the distribution of $\mathrm{DM_{host}}$,
the median value and variance are $e^{\mu_{\mathrm{host}}}$ and
$e^{\left(2\mu_{\mathrm{host}}+\sigma^{2}_{\rm host}\right)}[e^{\sigma^{2}_{\rm host}}-1]$, respectively.
Using the IllustrisTNG simulation, \cite{Zhang2020} derived the $\mathrm{DM_{host}}$ distributions
for repeating and nonrepeating FRBs at different redshifts. Following \cite{Zhang2020}, the localized
FRBs are divided into three types based on their host properties: repeating FRBs like FRB 20121102A (type I),
repeating FRBs like FRB 20180916B (type II), and nonrepeating FRBs (type III). For each host type,
\cite{Zhang2020} found that the median value of $\mathrm{DM_{host}}$ (i.e., $e^{\mu_{\mathrm{host}}}$)
increases with redshift as $e^{\mu_{\mathrm{host}}(z)}=\kappa(1+z)^{\alpha}$, where the best-fit values
of $\kappa$ and $\alpha$ are given in their work. In our calculation, we use their results to get
$\mu_{\mathrm{host}}$ and $\sigma_{\mathrm{host}}$ at any redshift of a localized FRB.

In theory, the average value of the IGM portion of DM is related to the cosmological distance scale
through \citep{2003ApJ...598L..79I,2004MNRAS.348..999I}
\begin{equation}\label{eq:DMIGM}
	\left\langle\mathrm{DM_{IGM}}(z)\right\rangle=\frac{3 c H_{0} \Omega_{b} f_{\mathrm{IGM}}}{8 \pi G m_{p}} \int_{0}^{z} \frac{(1+z^{\prime}) \chi_{e}(z^{\prime})}{E(z^{\prime})} d z^{\prime}\;,
\end{equation}
where $m_p$ is the proton mass, $H_0$ is the Hubble constant, $\Omega_{b}$ is the baryon density parameter,
$f_{\mathrm{IGM}}\simeq0.83$ is the fraction of baryons in the IGM \citep{1998ApJ...503..518F}, and
$E(z)$ is the dimensionless Hubble parameter that is dependent upon a given cosmological model.
The number of free electrons per baryon is $\chi_{e}(z)=\frac{3}{4} \chi_{e,\mathrm{H}}(z)+
\frac{1}{8}\chi_{e, \mathrm{He}}(z)$, where $\chi_{e,\mathrm{H}}(z)$ and $\chi_{e,\mathrm{He}}(z)$
denote, respectively, the ionization fractions of hydrogen and helium. At redshifts $z\lesssim3$
hydrogen and helium are fully ionized \citep{Meiksin2009,Becker2011}, allowing one to set
$\chi_{e,\mathrm{H}}= \chi_{e,\mathrm{He}}=1$, which therefore gives $\chi_{e}=7/8$.

It should be emphasized that the actual value of $\mathrm{DM_{IGM}}$ would vary significantly
around the mean $\langle\mathrm{DM_{IGM}}\rangle$ due to the large IGM fluctuations.
The probability distribution of $\mathrm{DM_{IGM}}$ can be derived from numerical simulations of
the IGM and galaxy haloes \citep{McQuinn2014,Prochaska2019a}, which can be well fitted by the analytic form
\citep{Macquart2020,Zhang2021}
\begin{equation}\label{eq:PIGM}
	P_{\mathrm{IGM}}(\Delta)=A \Delta^{-\beta} \exp \left[-\frac{\left(\Delta^{-\alpha}-C_{0}\right)^2}{2 \alpha^{2} \sigma_{\mathrm{IGM}}^{2}}\right],\;\;\; \Delta>0\;,
\end{equation}
where $\Delta\equiv\mathrm{DM}_{\mathrm{IGM}}/\left\langle\mathrm{DM}_{\mathrm{IGM}}\right\rangle$,
$A$ is a normalization constant, $\sigma_{\mathrm{IGM}}$ is an effective standard deviation,
$C_{0}$ is chosen such that the mean of the distribution is unity, and $\alpha$ and $\beta$
are two indices related to the inner density profile of gas in halos. Following \cite{Macquart2020},
we take $\alpha=3$ and $\beta=3$. Using the IllustrisTNG simulation, \cite{Zhang2021} derived
realistic distributions of $\mathrm{DM_{IGM}}$ at different redshifts. Here we adopt the best-fit
values of $A$, $C_0$, and $\sigma_{\mathrm{IGM}}$ for the $\mathrm{DM_{IGM}}$ distribution at different
redshifts corresponding to their results.

\begin{figure}
\vskip-0.1in
    \includegraphics[width=\columnwidth]{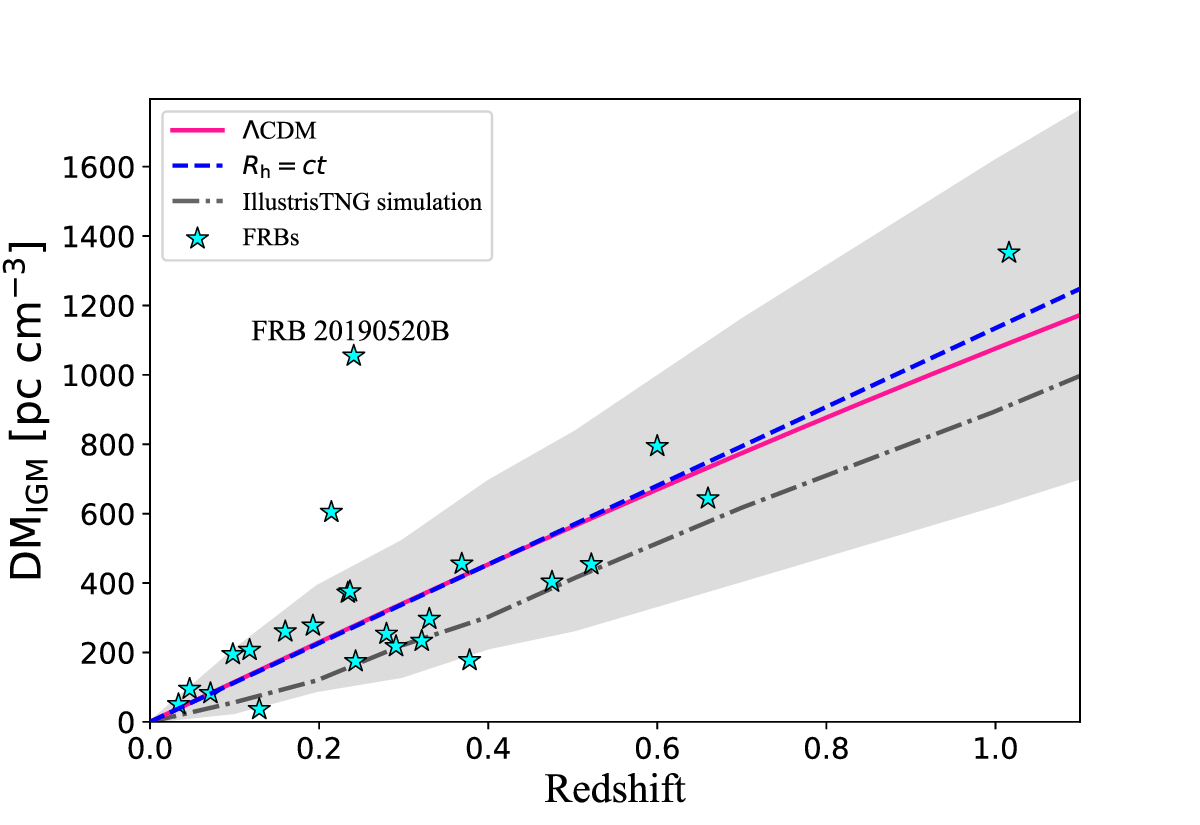}
    \caption{DM--redshift correlation for FRBs. Data points are estimations
    of the IGM portion $\mathrm{DM_{IGM}}$ versus redshift for 24 localized FRBs.
    The $\mathrm{DM_{IGM}}$ values are derived from the observed $\mathrm{DM_{obs}}$
    after removing the contributions from our Galaxy and the host galaxy.
    The $\mathrm{DM_{ISM}^{MW}}$ term is estimated from the Galactic ISM model,
    and $\mathrm{DM_{halo}^{MW}}$ and $\mathrm{DM_{host}}$ are assumed to be
    $50\;\mathrm{pc\;cm^{-3}}$ and $50(1+z)^{-1}\;\mathrm{pc\;cm^{-3}}$, respectively.
    The dot-dashed line represents the $\mathrm{DM_{IGM}}$ result deduced from
    the IllustrisTNG simulation and the gray shaded region encloses the 95\%
    confidence region \citep{Zhang2021}. The other two lines correspond to
    the predictions of different cosmological models: (solid)
    flat $\Lambda$CDM, with $H_0=95.8$ $\mathrm{km\;s^{-1}\;Mpc^{-1}}$ and
    $\Omega_{\rm m}=0.66$; (dashed) the $R_{\rm h}=ct$ cosmology, with $H_0=94.2$
    $\mathrm{km\;s^{-1}\;Mpc^{-1}}$.}
    \label{fig1}
\end{figure}

As of today, 26 FRBs have been localized to a host galaxy\footnote{Here the terminology of ``localized''
means the host galaxy has been identified. However, ``localized'' can also mean that
the radio localization is of order arcsec, and different groups have different definitions
of how well-localized an FRB has to be to call it ``localized'' (usually the cutoff is
the same as their instrument's angular resolution). This distinction is important in the context of
this work because a given ``localization'' accuracy (i.e. angular error) is more likely to
lead to the identification of the host galaxy for a near-Universe FRB.},
including 18 apparently non-repeaters and 8 repeaters.\footnote{Note that
there might be a potential bias from including localized repeating FRBs that have only been
localized because they repeat. But as long as the localization is as reliable as any of
the others, repeating sources should be fine. Moreover, since the current localized sample
size is admittedly small, we use all the FRBs for our analysis.}
Table~\ref{tab1} lists the redshifts, $\mathrm{DM_{obs}}$ (with corresponding uncertainties
$\sigma_{\rm obs}$), and $\mathrm{DM_{ISM}^{MW}}$ of all 26 localized FRBs.
FRB 20200110E is located in a globular cluster in the nearby galaxy M81,
whose distance is only 3.6 Mpc \citep{Bhardwaj2021b,Kirsten2022}. For such a short distance,
the peculiar velocity dominates over the Hubble flow so that it has a negative spectroscopic
redshift $z=-0.0001$. Since the $\mathrm{DM_{obs}}$ value of FRB 20181030A is only
$103.5\;\mathrm{pc\;cm^{-3}}$ \citep{2021ApJ...919L..24B}, its extragalactic $\mathrm{DM_{ext}}$
will be reduced to a negative value after removing $\mathrm{DM_{ISM}^{MW}}$ and $\mathrm{DM_{halo}^{MW}}$.
That is, the integral upper limit ($\mathrm{DM_{ext}}\equiv\mathrm{DM_{obs}-DM_{ISM}^{MW}-DM_{halo}^{MW}}$)
in the probability of the extragalactic $\mathrm{DM_{ext}}$ (see Eq.~\ref{eq:P_FRB}) will become negative.
Thus, neither FRB 20200110E nor FRB 20181030A are available for our subsequent maximum likelihood analysis,
so we exclude them from the localized FRB sample. Figure~\ref{fig1} shows the estimated
$\mathrm{DM_{IGM}}$ and measured $z$ values for the remaining 24 FRBs. We have estimated $\mathrm{DM_{IGM}}$
by deducting the following contributions from the measured $\mathrm{DM_{obs}}$ value: $\mathrm{DM_{ISM}^{MW}}$
based on the Galactic ISM model; and the assumed $\mathrm{DM_{halo}^{MW}}=50\;\mathrm{pc\;cm^{-3}}$
and $\mathrm{DM_{host}}=50(1+z)^{-1}\;\mathrm{pc\;cm^{-3}}$. In Figure~\ref{fig1}, the dot-dashed line is
the $\mathrm{DM_{IGM}}$ result estimated from the IllustrisTNG simulation with 95\% confidence region
\citep{Zhang2021}. FRB 20190520B is associated with a dwarf host galaxy at $z=0.241$.
It was found that FRB 20190520B deviates significantly from the general trend of the
$\mathrm{DM_{IGM}}$--$z$ relation, with an unprecedented DM contribution from its host galaxy \citep{Niu2022}.
Since $\mathrm{DM_{host}}$ is assumed to follow a lognormal distribution with
an asymmetric tail to large values (see Eq.~\ref{eq:Phost}), our model ought to be able to
account for FRB 20190520B well.
Therefore, we do not exclude this FRB from our analysis. In total, there are 24 FRBs for us to
consider in the rest of this paper.

\section{Model Selection}\label{sec:model}
We are now in position to use the 24 DM--$z$ measurements of the FRB sample
to test and compare certain cosmological models. The cosmological parameters are
optimized via a maximization of the joint likelihood function \citep{Macquart2020}:
\begin{equation}
	\mathcal{L}=\prod_{i=1}^{24} P_{i}\left(\mathrm{DM}_{\mathrm{ext}, i}\right)\;,
\end{equation}
where $P_{i}(\mathrm{DM}_{\mathrm{ext}, i})$ is the probability of individual observed $\mathrm{DM_{obs}}$
corrected for our Galaxy, i.e., $\mathrm{DM_{ext}}\equiv\mathrm{DM_{obs}}-\mathrm{DM_{ISM}^{MW}}
-\mathrm{DM_{halo}^{MW}}=\mathrm{DM_{IGM}}+\mathrm{DM_{host}}$. For an FRB at redshift $z_i$,
we have
\begin{equation}\label{eq:P_FRB}
	\begin{split}
	P_{i}\left(\mathrm{DM}_{\mathrm{ext}, i} \right) &=
	\int_{0}^{\mathrm{DM}_{\mathrm{ext}, i}}
        P_{\text {host}}\left(\mathrm{DM}_{\mathrm {host}}\right)\times \\
	&P_{\mathrm {IGM}}\left(\mathrm{DM}_{\mathrm{ext}, i}-\mathrm{DM_{host}}\right) \mathrm{dDM}_{\mathrm {host}}\;,
	\end{split}
\end{equation}
where $P_{\text {host}}(\mathrm{DM_{host}})$ and $P_{\text {IGM}}(\mathrm{DM_{IGM}})$ (see
Eqs.~\ref{eq:Phost} and \ref{eq:PIGM}) are the probability density functions for
$\mathrm{DM_{host}}$ and $\mathrm{DM_{IGM}}$, respectively.
To determine the model predictions for the average value of
$\left\langle\mathrm{DM_{IGM}}(z)\right\rangle$ in Equation~(\ref{eq:DMIGM}), we need
an expression for the model-dependent dimensionless expansion rate $E(z)$. Here we discuss
how the fits are obtained for $\Lambda$CDM and the $R_{\rm h}=ct$ universe.
It is worth emphasizing that the IGM baryon fraction $f_{\mathrm{IGM}}$
is fixed to be 0.83 \citep{1998ApJ...503..518F}. Moreover, there is a degeneracy
between the Hubble constant $H_0$ and the baryon density parameter $\Omega_{b}$
in Equation~(\ref{eq:DMIGM}), so we adopt a fixed value $\Omega_{b}=0.0493$
for the sake of constraining $H_0$ \citep{2020A&A...641A...6P}.

\begin{figure}
\centerline{
	\includegraphics[width=\columnwidth]{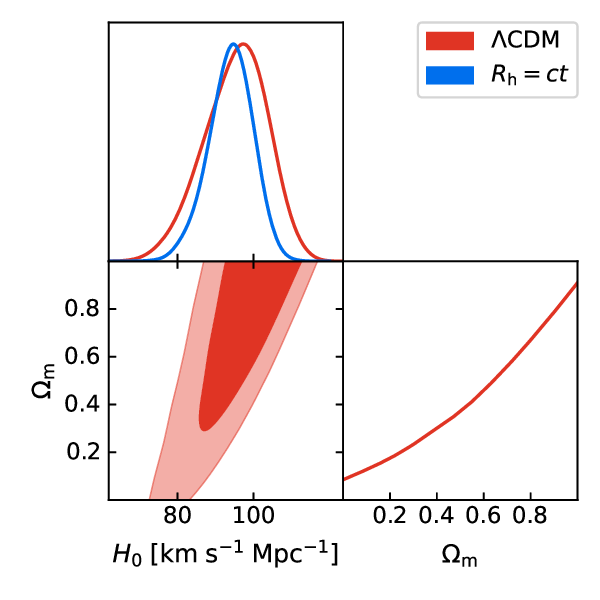}}
\vskip-0.1in
    \caption{1D marginalized posterior distributions and 2D $1-2\sigma$ contour regions
representing the cosmological parameters $H_0$ and/or $\Omega_{\rm m}$, constrained by
24 localized FRBs. The colours represent: flat $\Lambda$CDM (red contours) and $R_{\rm h}=ct$ (blue line).}
    \label{fig2}
\end{figure}

In flat $\Lambda$CDM, the dimensionless Hubble parameter ($H[z]/H_0$) is well approximated by
\begin{equation}
E^{\Lambda {\rm CDM}}\left(z\right)=\left[\Omega_{\rm m}\left(1+z\right)^{3}+
\Omega_{\Lambda}\right]^{1/2}\;,
\end{equation}
where $\Omega_{\rm m}$ is the present-day scaled matter energy density and $\Omega_{\Lambda}
=1-\Omega_{\rm m}$ is the vacuum energy density (assuming spatial flatness). We
allow $\Omega_{\rm m}$ to vary. In this model, the free parameters are thus $H_0$ and
$\Omega_{\rm m}$.

On the other hand, the dimensionless expansion rate in the $R_{\rm h}=ct$ universe
\citep{2007MNRAS.382.1917M,2012MNRAS.419.2579M,Melia2020}, is given as
\begin{equation}
E^{R_{\rm h}=ct}\left(z\right)=1+z\;.
\end{equation}
In this model, we have only one free parameter that need to be optimized: $H_0$.

For each model, we use the Python Markov chain Monte Carlo (MCMC) module, EMCEE
\citep{Foreman2013}, to explore the posterior probability distributions of the free
parameters. In our MCMC analysis, we set wide flat priors on $H_0\in[0,\;150]$
$\mathrm{km\;s^{-1}\;Mpc^{-1}}$ and/or $\Omega_{\rm m}\in[0,\;1]$.
The 1D marginalized posterior distributions and 2D plots of the $1-2\sigma$
confidence regions for these parameters, constrained by 24 localized FRBs,
are displayed in Figure~\ref{fig2}. For $\Lambda$CDM, these red contours show that
at the $1\sigma$ confidence level, the inferred value is $H_0=95.8^{+7.8}_{-9.2}$
$\mathrm{km\;s^{-1}\;Mpc^{-1}}$, but that $\Omega_{\rm m}$ is poorly constrained.
Because the prior on $\Omega_{\rm m}$ is uniform over the interval $[0,\;1]$, only a median
value of $\sim0.66$ can be estimated.
The maximum value of the likelihood function for optimized flat $\Lambda$CDM is given by
$-2\ln \mathcal{L}=308.97$, which will be used when comparing models using information criteria.

Given the high uncertainty in $\mathrm{DM_{halo}^{MW}}$, here we set
a wide flat prior $\mathrm{DM_{halo}^{MW}}\in[5,\;80]$ $\mathrm{pc\;cm^{-3}}$.
Taking a larger value of $\mathrm{DM_{halo}^{MW}}$ will lead to a smaller value of $\mathrm{DM_{IGM}}$,
thus yielding a lower value of $H_0$. Based on the estimate of \cite{Prochaska2019a},
we next consider whether a narrower flat prior $[50,\;80]$ $\mathrm{pc\;cm^{-3}}$ affects
the $H_0$ measurement. For $\Lambda$CDM, we have also performed a parallel comparative analysis
of the FRB data by marginalizing $\mathrm{DM_{halo}^{MW}}$ over a flat prior of 50--80 $\mathrm{pc\;cm^{-3}}$.
In this case, the constraints on the cosmological parameters are
$H_0=77.1^{+7.0}_{-7.4}$ $\mathrm{km\;s^{-1}\;Mpc^{-1}}$ and $\Omega_{\rm m}\sim0.56$.
As expected, the adoption of a higher value of $\mathrm{DM_{halo}^{MW}}$ leads to a lower $H_0$ measurement.
It is obvious that the $\mathrm{DM_{halo}^{MW}}$ prior has a significant impact on the $H_0$ measurement.

Even though the large $H_0$ ($95.8^{+7.8}_{-9.2}$ $\mathrm{km\;s^{-1}\;Mpc^{-1}}$) for the given flat prior of 5--80
$\mathrm{pc\;cm^{-3}}$ is the result we obtain with the current best knowledge of
$\mathrm{DM_{halo}^{MW}}$, we want to point out some potential caveats to this measurement.
(i) In general, an underestimated host contribution could be the most likely cause of an elevated
$H_0$. The probability density function for $\mathrm{DM_{host}}$ has limited theoretical
motivation. We don't know if $P_{\text {host}}(\mathrm{DM_{host}})$ can represent the shape
of the host contribution well, and as we have already noticed, there are bursts like FRB 20190520B
with a large $\mathrm{DM_{host}}$ contribution. \hbox{(ii)} The effect of observational biases
against high-DM FRBs might be the other cause of an elevated $H_0$.
Here we use the DM--$z$ relation of localized FRBs to constrain $H_0$. However, at very high DMs,
only nearby FRBs are detectable, since an event must be observed with sufficiently high fluence to overcome the
detection bias against high DM. That is, observational biases cause the DM--$z$ relation to become inverted
\citep{2022MNRAS.509.4775J}. More localized FRBs will be necessary to quantitatively estimate this effect.

Given the wide flat prior $\mathrm{DM_{halo}^{MW}}\in[5,\;80]$ $\mathrm{pc\;cm^{-3}}$,
the results of fitting the 24 $\mathrm{DM_{obs}}$--redshift data with the $R_{\rm h}=ct$
universe are shown as the blue line in Figure~\ref{fig2}. We have $H_0=94.2^{+5.6}_{-6.2}$
$\mathrm{km\;s^{-1}\;Mpc^{-1}}$. For the optimized $R_{\rm h}=ct$ fit, the maximum value of
the likelihood function yields $-2\ln \mathcal{L}=310.18$.

To assess how these two models compare with each other, we show in
Figure~\ref{fig1} the expected relations between $\mathrm{DM_{IGM}}$ and redshift for
$\Lambda$CDM (with the best-fitting parameters $H_0=95.8$ $\mathrm{km\;s^{-1}\;Mpc^{-1}}$
and $\Omega_{\rm m}=0.66$; solid line) and for $R_{\rm h}=ct$
(with $H_0=94.2$ $\mathrm{km\;s^{-1}\;Mpc^{-1}}$; dashed line). One can see from
this plot that the differences between alternative cosmological models emerge more
prominently at higher redshifts.

Since these two models have different numbers of free parameters, however, a comparison
of the likelihoods judging which cosmology is a better match to the data must be based on
model selection tools. One tool commonly used to differentiate between cosmological models
is the Bayesian Information Criterion (BIC; \citealt{1978AnSta...6..461S}),
\begin{equation}
\mathrm{BIC}=-2\ln \mathcal{L}+\left(\ln N\right)f\;,
\end{equation}
where $N$ is the number of data points (here 24) and $f$ is the number of free parameters
(four for $\Lambda$CDM and three for $R_{\rm h}=ct$). With ${\rm BIC}_{\alpha}$ characterizing
model $\mathcal{M}_{\alpha}$, the unnormalized confidence that this model is true is the
Bayesian weight $\exp(-{\rm BIC}_{\alpha}/2)$. The likelihood of model $\mathcal{M}_{\alpha}$
being correct relative to the other one is then
\begin{equation}\label{eq:Palpha}
P(\mathcal{M}_{\alpha})=\frac{\exp(-{\rm BIC_\alpha}/2)}{\exp(-{\rm BIC_{1}}/2)+\exp(-{\rm BIC_{2}}/2)}\;.
\end{equation}
The model with the smaller BIC score is evidently preferred by this criterion. Our analysis
shows that $R_{\rm h}=ct$ is favored over $\Lambda$CDM with a likelihood of $\sim73\%$ versus $\sim27\%$.

\section{Monte Carlo Simulations}\label{sec:MC}

Though $R_{\rm h}=ct$ is somewhat favored by this analysis, it is obvious that the
current sample of FRBs is too small to differentiate strongly between this model and standard
$\Lambda$CDM. To forecast how many localized FRBs are required to really distinguish the models,
we shall here produce mock samples of FRBs through Monte Carlo simulations. In using the BIC
model selection tool, the difference $\Delta{\rm BIC}\equiv$ BIC$_1-$ BIC$_2$ determines the
extent to which model $\mathcal{M}_{1}$ is favored over $\mathcal{M}_{2}$. The outcome of the
difference is judged `positive' in the range $\Delta{\rm BIC}=2-6$, `strong' for
$\Delta{\rm BIC}=6-10$, and `very strong' for $\Delta{\rm BIC}>10$. In this section, we shall
provide a quantitative assessment of how many FRB detections are necessary to rule out
$R_{\rm h}=ct$ or $\Lambda$CDM, by conservatively seeking an outcome even beyond
$\Delta{\rm BIC}\simeq11.62$. That is, we shall consider two cases: one in which the background
cosmological model is assumed to be $\Lambda$CDM, and the other in which it is $R_{\rm h}=ct$, and
we shall estimate the number of FRBs required in each case to rule out the alternative model with
a likelihood $\sim 99.7\%$ versus $\sim 0.3\%$, corresponding to a $3\sigma$ confidence level.

\begin{figure}
	\includegraphics[width=\columnwidth]{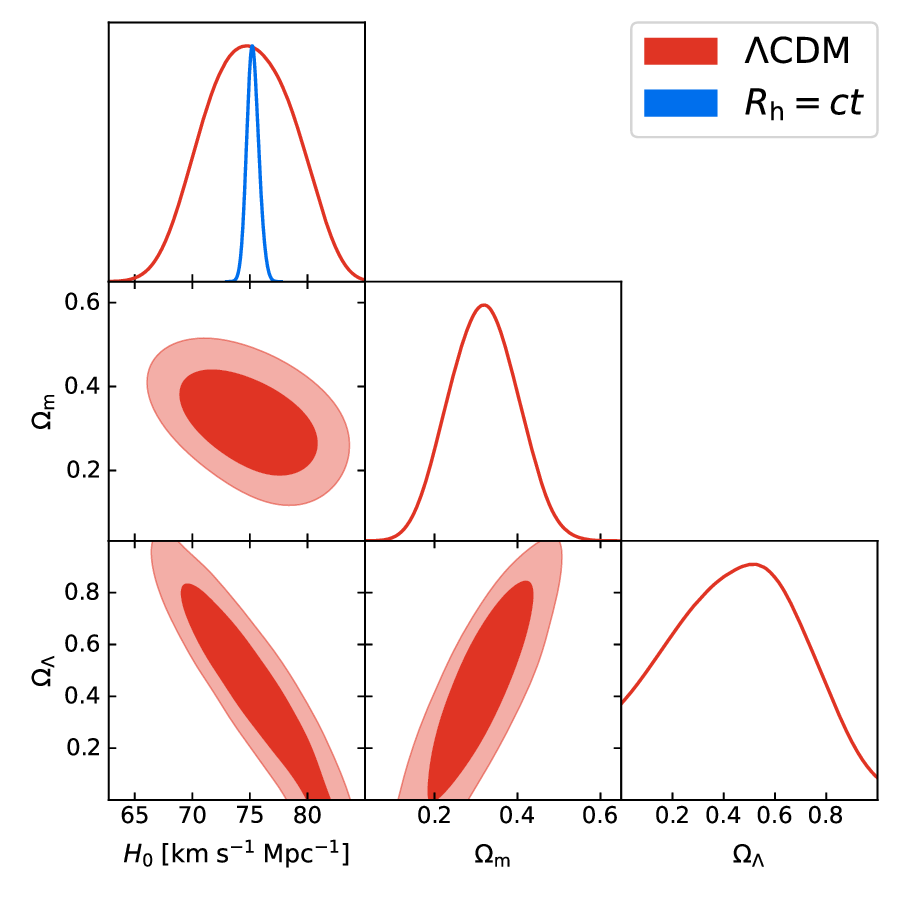}
\vskip-0.1in
    \caption{1D marginalized posterior distributions and 2D $1-2\sigma$ contour regions
for the cosmological parameters $H_0$, $\Omega_{\rm m}$, and/or $\Omega_{\Lambda}$,
using the simulated sample with 1,150 FRBs, assuming $\Lambda$CDM as the background cosmology.
The colours represent: the best-fit non-flat $\Lambda$CDM (red contours) and $R_{\rm h}=ct$ (blue line).}
    \label{fig3}
\end{figure}

The simulated FRBs are each characterized by the parameters ($z$, $\mathrm{DM_{ext}}$).
The procedures of simulation are as follows:
\begin{enumerate}
  \item We assume that FRBs and gamma-ray bursts have the similar redshift distribution
   \citep{2011ApJ...738...19S,Zhou2014}, which is estimated as $P(z)\propto ze^{-z}$.
   The redshift $z$ of our synthetic FRBs is generated randomly from this distribution
   in the redshift range $0 < z < 3$.
   \item With the mock $z$, $\mathrm{DM_{host}}$
  is generated randomly from the probability distribution in Equation~(\ref{eq:Phost}).
  \item With the mock $z$, we infer the mean value $\langle\mathrm{DM_{IGM}}(z)\rangle$
  from Equation~(\ref{eq:DMIGM}) corresponding either to a flat $\Lambda$CDM cosmology
  with $\Omega_{\rm m}=0.315$ and $H_{0}=67.4$ $\mathrm{km\;s^{-1}\;Mpc^{-1}}$
  (Section~\ref{subsec:LCDM}), or the $R_{\rm h}=ct$ universe with $H_{0}=67.4$
  $\mathrm{km\;s^{-1}\;Mpc^{-1}}$ (Section~\ref{subsec:Rh}).
  \item With the selected $z$ and $\langle\mathrm{DM_{IGM}}(z)\rangle$, $\mathrm{DM_{IGM}}$
  is generated randomly from the probability distribution in Equation~(\ref{eq:PIGM}).
  \item With the mock $\mathrm{DM_{host}}$ and $\mathrm{DM_{IGM}}$, we calculate $\mathrm{DM_{ext}}$
  according to Equation~(\ref{eq:DMext}).
 \end{enumerate}

For a mock FRB sample with a fixed number $N$, we optimize the model fits using the analysis method
described in the previous section.
To ensure the final results are unbiased, we repeat this process 100 times for a fixed number $N$ of
FRBs by using different noise seeds. The mock FRB sample is enlarged until the likelihood criterion
discussed above is reached.

\begin{figure}
	\includegraphics[width=\columnwidth]{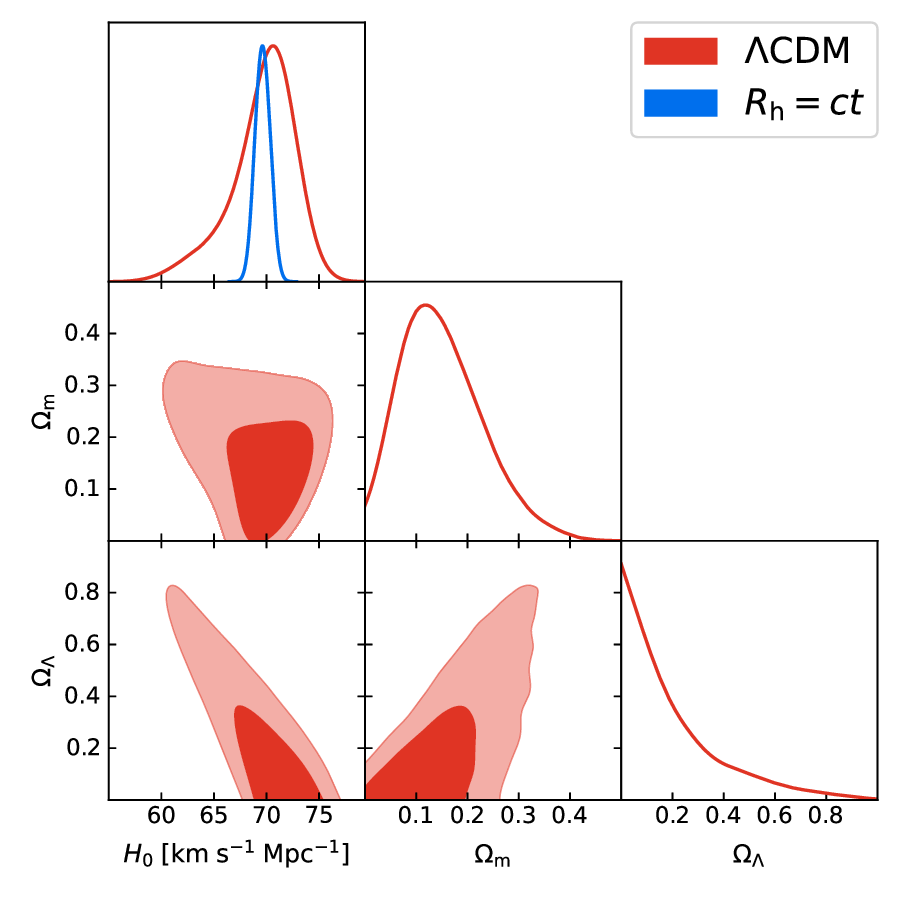}
\vskip-0.1in
    \caption{Same as Figure~\ref{fig3}, except now using the simulated sample with
    550 FRBs, assuming $R_{\rm h}=ct$ as the background cosmology.}
    \label{fig4}
\end{figure}

\subsection{Assuming $\Lambda$CDM as the Background Cosmology}
\label{subsec:LCDM}
In the first case, we assume that the background cosmological model is $\Lambda$CDM,
and seek the minimum sample size required to rule out $R_{\rm h}=ct$ at the $3\sigma$
confidence level. We have found that a sample of at least 1,150 FRBs is necessary to
achieve this goal. The best-fit parameters corresponding to
the $\Lambda$CDM model for these simulated FRBs are displayed as red contours
in Figure~\ref{fig3}. To allow for greater flexibility in this fit, we relax
the assumption on flatness, and allow $\Omega_{\Lambda}$ to be a free parameter, along with
$\Omega_{\rm m}$. In non-flat $\Lambda$CDM, the dimensionless Hubble parameter is given by
$E^{\Lambda {\rm CDM}}\left(z\right)=\left[\Omega_{\rm m}\left(1+z\right)^{3}+
\Omega_{\Lambda}+\Omega_{k}\left(1+z\right)^{2}\right]^{1/2}$, where
$\Omega_{k}=1-\Omega_{\rm m}-\Omega_{\Lambda}$ is the spatial curvature of the Universe.
The best-fit values for $\Lambda$CDM using the simulated sample with 1,150 FRBs
are $H_0=74.8^{+4.1}_{-3.9}$ $\mathrm{km\;s^{-1}\;Mpc^{-1}}$, $\Omega_{\rm m}=0.32^{+0.08}_{-0.08}$,
and $\Omega_{\Lambda}=0.46^{+0.25}_{-0.27}$.

In Figure~\ref{fig3}, we also show the corresponding 1D posterior distribution of $H_0$
for the $R_{\rm h}=ct$ universe (see the blue line). The best-fit value for the simulated
sample is $H_0=75.2^{+0.5}_{-0.5}$ $\mathrm{km\;s^{-1}\;Mpc^{-1}}$. With $N=1,150$,
our analysis shows that the BIC would favor non-flat $\Lambda$CDM
over $R_{\rm h}=ct$ by an overwhelming likelihood of $99.7\%$ versus only $0.3\%$
(i.e., the prescribed $3\sigma$ confidence level).

\subsection{Assuming $R_{\rm h}=ct$ as the Background Cosmology}
\label{subsec:Rh}
Assuming $R_{\rm h}=ct$ as the background cosmology, we have found that a minimum of 550 FRBs
are required to rule out the non-flat $\Lambda$CDM model. The 1D marginalized posterior distributions
and 2D plots of the $1-2\sigma$ confidence regions for two-parameter combinations, constrained by
550 simulated FRBs are displayed in Figure~\ref{fig4}. For non-flat $\Lambda$CDM,
these red contours show that at the $1\sigma$ confidence level, the inferred values are
$H_0=70.1^{+2.4}_{-3.4}$ $\mathrm{km\;s^{-1}\;Mpc^{-1}}$ and $\Omega_{\rm m}=0.14^{+0.09}_{-0.07}$,
but that $\Omega_{\Lambda}$ is poorly constrained; only an upper limits of $<0.24$ can be estimated.

The corresponding 1D posterior distribution of $H_0$ for the $R_{\rm h}=ct$ universe
is shown as the blue line in Figure~~\ref{fig4}. The best-fit value for the simulated sample
with 550 FRBs is $H_0=69.6^{+0.7}_{-0.7}$ $\mathrm{km\;s^{-1}\;Mpc^{-1}}$.
With $N=550$, our analysis shows that in this case the BIC
would favor $R_{\rm h}=ct$ over non-flat $\Lambda$CDM by an overwhelming likelihood of $99.7\%$
versus only $0.3\%$ (i.e., the prescribed $3\sigma$ confidence level).

\section{Investigating the Hubble Tension}\label{sec:HT}
The $\mathrm{DM_{IGM}}$--$z$ relation of FRBs has been used to probe cosmology, but
this approach is limited by the poorly constrained dispersion measure, $\mathrm{DM_{host}}$,
contributed by the host galaxy \citep{Macquart2020}. Most of the previous work has been
carried out by assuming a significant positive skew for the distribution of $\mathrm{DM_{host}}$
and/or $\mathrm{DM_{IGM}}$, including the treatment described in Section~\ref{sec:model}.
Here, we present a straightforward method of deriving an upper limit to $H_0$ without having
to make any assumption regarding the probability distribution for $\mathrm{DM_{host}}$.

Equation~(\ref{eq:DMIGM}) shows that the theoretical $\mathrm{DM_{IGM}}$ at any given redshift
is proportional to $H_0$. An upper limit for $H_0$ can therefore be obtained by requiring that
$\mathrm{DM_{ext}}$ (derived from Eq.~\ref{eq:DMext}) be at least as large as
$\mathrm{DM_{IGM}}$ at the corresponding redshifts. In other words, $\mathrm{DM_{ext}}$
can be used to derive an upper limit to $\mathrm{DM_{IGM}}$, and thus an average upper
limit\footnote{Due to the large fluctuations in density, $\mathrm{DM_{IGM}}$ at a given redshift
can vary by as much as 40\% along different lines of sight at $z\sim1$ \citep{McQuinn2014}.
In principle, the derived `upper limit' may thus be exceeded for individual FRBs.} for $H_0$,
without the inclusion of the poorly known $\mathrm{DM_{host}}$. In addition,
$\mathrm{DM_{halo}^{MW}}$ in Equation~(\ref{eq:DMext}) is also uncertain. Its inclusion
would yield a more stringent upper limit on $H_0$. To be as conservative as possible,
we therefore avoid introducing $\mathrm{DM_{halo}^{MW}}$ in Equation~(\ref{eq:DMext}).
Hereafter, we thus conservatively constrain $H_0$ using the less restrictive extragalactic
dispersion measures $\mathrm{DM'_{ext}}=\mathrm{DM_{obs}}-\mathrm{DM_{ISM}^{MW}}$.
The method of placing an upper bound on $H_0$ was first developed by
\cite{2022MNRAS.516.4862J}, though those authors fixed $\Omega_{b}h^{2}$ rather than
$\Omega_{b}$, thus producing a `lower' bound. This approach results in $\mathrm{DM_{ext}}$
becoming negatively correlated with $H_0$, hence producing a lower limit for $H_0$
rather than an upper limit.

\begin{figure}
	\includegraphics[width=\columnwidth]{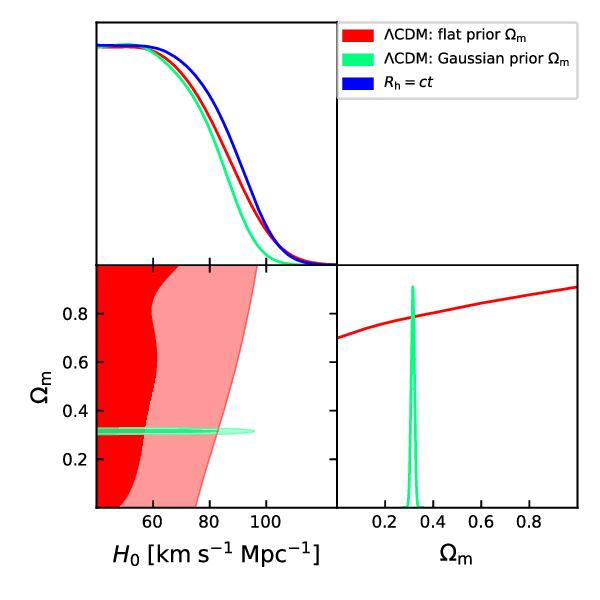}
\vskip-0.2in
    \caption{1D marginalized posterior distributions and 2D $1-2\sigma$ contour regions
for $H_0$ and/or $\Omega_{\rm m}$. The colours represent: $\Lambda$CDM with a flat prior
$\Omega_{\rm m}\in[0,\;1]$ (red), $\Lambda$CDM with a Gaussian prior $\Omega_{\rm m}=0.315\pm0.007$
(green), and $R_{\rm h}=ct$ (blue).}
    \label{fig5}
\end{figure}

Utilizing the observed data $\mathbf{D}$ (with the `extragalactic' dispersion measures
at redshifts $z_i$ taken to be $\mathrm{DM'_{ext,\emph{i}}}$) and some prior knowledge
about the hypothetical cosmological models (for which the parameters are denoted by
the vector $\boldsymbol{\theta}$), the posterior distribution of the free parameters
is modeled through the half-Gaussian \hbox{(log-)}likelihood
\citep{2022JHEAp..36...27V,2022ApJ...928..165W},
\begin{equation}\label{eq:halfGaussian}
\ln{\mathcal L}\left(\boldsymbol{\theta}\mid\mathbf{D}\right) = -\frac{1}{2}\sum_{i}^{24} \left\lbrace \begin{array}{ll}
\delta_{i}^{2}\left(\boldsymbol{\theta}\right)/\sigma_{\mathrm{tot}, i}^{2}~~~~~~{\rm if}~~\delta_{i}\left(\boldsymbol{\theta}\right)<0\\
0~~~~~~~~~~~~~~~~~~~~~~{\rm if}~~\delta_{i}\left(\boldsymbol{\theta}\right)\geq0\;,\\
\end{array} \right.
\end{equation}
where $\delta_{i}\equiv\mathrm{DM'_{ext,\emph{i}}}-\left\langle\mathrm{DM_{IGM}}(\boldsymbol{\theta},\;z_{i})\right\rangle$
and $\sigma_{\mathrm{tot}, i}^{2}=\sigma_{\mathrm{obs}, i}^{2}+\sigma_{\mathrm{ISM}}^{2}
+\sigma_{\mathrm{IGM}}^{2}(z_i)$. Here $\sigma_{\mathrm{obs}}$, $\sigma_{\mathrm{ISM}}$,
and $\sigma_{\mathrm{IGM}}$ are the uncertainties associated with $\mathrm{DM_{obs}}$,
$\mathrm{DM_{ISM}^{MW}}$, and the mean $\mathrm{DM_{IGM}}(z)$, respectively. The Australia Telescope
National Facility Pulsar Catalogue indicates that the average uncertainty in $\mathrm{DM_{ISM}^{MW}}$
for sources at high Galactic latitude is about 10 $\mathrm{pc\;cm^{-3}}$ \citep{2005AJ....129.1993M},
and we thus adopt this value for $\sigma_{\mathrm{ISM}}$.
Density fluctuations in the large-scale structure could lead to
substantial sightline-to-sightline scatter around the mean $\mathrm{DM_{IGM}}(z)$.
Here we associate the standard deviation $\sigma_{\mathrm{IGM}}(z)$ that results from
the numerical simulation of \cite{McQuinn2014} to $\mathrm{DM_{IGM}}(z)$ (see their Fig.~1).
Equation~(\ref{eq:halfGaussian}) expresses the fact that:
$a)$ since $\mathrm{DM'_{ext}}$ must not be smaller than the IGM portion, parameters
for which $\mathrm{DM'_{ext}}$ is smaller than $\mathrm{DM_{IGM}}$ (i.e., $\delta_{i}
(\boldsymbol{\theta})<0$) are exponentially unlikely, and this means the fit worsens
as $\mathrm{DM'_{ext}}$ decreases relative to $\mathrm{DM_{IGM}}$; $b)$ when $\mathrm{DM'_{ext}}$
is larger than $\mathrm{DM_{IGM}}$ (i.e., $\delta_{i}(\boldsymbol{\theta})\geq0$),
the parameters are equally likely and cannot be differentiated on the basis of the FRB data alone.

For the flat $\Lambda$CDM model, the free parameters to be inferred in
Equation~(\ref{eq:halfGaussian}) are $\boldsymbol{\theta}=\{H_{0},\;\Omega_{\rm m}\}$,
for which we set flat priors: $H_0\in[0,\;150]$ $\mathrm{km\;s^{-1}\;Mpc^{-1}}$
and $\Omega_{\rm m}\in[0,\;1]$. The resulting constraints are displayed as red contours
in Figure~\ref{fig5}, which show that, whereas $\Omega_{\rm m}$ is poorly constrained,
$H_{0}$ is assigned an upper limit, as expected. At the 95\% confidence level,
we find that $H_{0}<89.0$ $\mathrm{km\;s^{-1}\;Mpc^{-1}}$ (the quoted upper limit
corresponds to the 95th percentile of the posterior distribution of $H_{0}$).
This upper limit is in good agreement with the latest local measurement
($73.04\pm1.04$ $\mathrm{km\;s^{-1}\;Mpc^{-1}}$; \citealt{2022ApJ...934L...7R}).

\begin{figure}
\vskip-0.1in
    \includegraphics[width=\columnwidth]{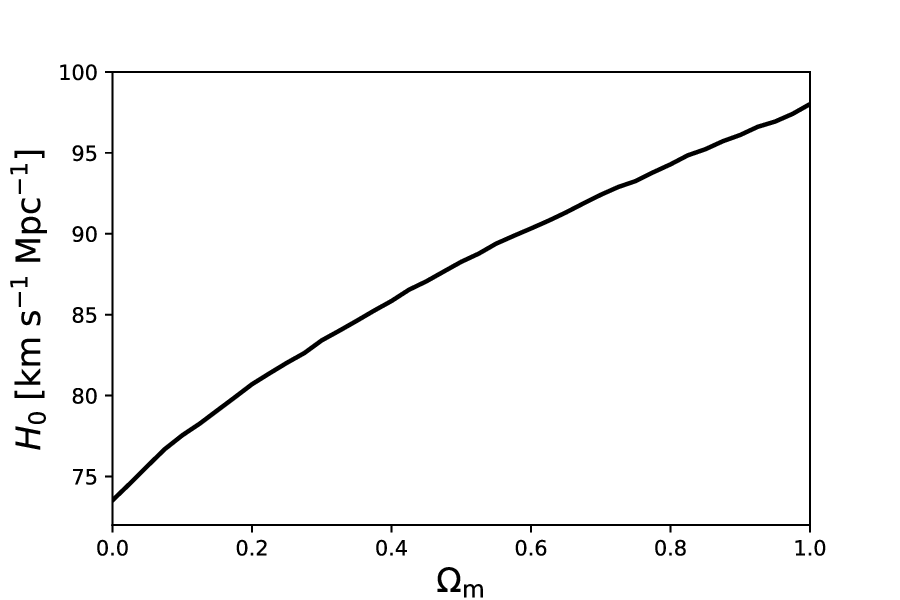}
    \caption{Inferred upper limit of $H_{0}$ as a function of $\Omega_{\rm m}$.}
    \label{fig6}
\end{figure}

We carry this analysis further by investigating how the limit on $H_0$ is affected by the
$\Omega_{\rm m}$ prior. First, we assume a Gaussian prior $\Omega_{\rm m}=0.315\pm0.007$,
informed by the {\it Planck} observations \citep{2020A&A...641A...6P}. The green contours in
Figure~\ref{fig5} show the joint $H_0$--$\Omega_{\rm m}$ posterior distributions obtained
with the flat prior on $H_0$ and this Gaussian prior on $\Omega_{\rm m}$. In this case,
the upper limit on the Hubble constant is $H_{0}<83.7$ $\mathrm{km\;s^{-1}\;Mpc^{-1}}$.
Next, we vary $\Omega_{\rm m}$ over the range $(0,1)$. The inferred upper limit to $H_{0}$
is shown as a function of $\Omega_{\rm m}$ in Figure~\ref{fig6}.

Clearly, the upper limit to $H_{0}$ is very sensitive to the value of $\Omega_{\rm m}$.
This echoes an interesting possibility proposed earlier (see, e.g., \citealt{Melia:2015})
that the disparity between the value of $H_0$ at high and low redshifts may in the end
be due to a different emphasis placed on $\Omega_{\rm m}$ by the CMB analysis versus
its relevance to the local region. To find a best fit with $\Lambda$CDM, we need to
optimize $H_0$ and $\Omega_{\rm m}$ together, and Figure~\ref{fig6} shows that these
parameters are tightly correlated. In other words, one could definitely argue that
new physics would be needed if $H_{0}$ were the only quantity changing between
high-$z$ and low-$z$ but, given this correlation, arising from a different emphasis
being placed on $\Omega_{\rm m}$, the changing $H_0$ may just be due to a biasing of
$\Omega_{\rm m}$ at either high or low redshifts.

In the $R_{\rm h}=ct$ cosmology, there is only one free parameter, i.e., $\boldsymbol{\theta}=\{H_{0}\}$.
With the same flat prior on $H_0$, we find an upper limit of $H_{0}<90.0$ $\mathrm{km\;s^{-1}\;Mpc^{-1}}$
at the 95\% confidence level (see the blue line in Figure~\ref{fig5}), consistent with the value inferred from
the local distance ladder. Again, the quoted upper limit corresponds to the 95th percentile
of the posterior distribution of $H_{0}$.

As shown in Section~\ref{sec:model}, if a wide flat prior, $\mathrm{DM_{halo}^{MW}}\in[5,\;80]$
$\mathrm{pc\;cm^{-3}}$, is considered, we have $H_0=95.8^{+7.8}_{-9.2}$ $\mathrm{km\;s^{-1}\;Mpc^{-1}}$
in our best fit analysis for $\Lambda$CDM. But in the analysis of deriving an upper limit
to $H_0$, we have $H_{0}<89.0$ $\mathrm{km\;s^{-1}\;Mpc^{-1}}$ for $\Lambda$CDM.
It seems strange that the best fit for $\Lambda$CDM requires a Hubble constant
greater than the upper limit established at the 95\% confidence level. This oddity is
due to the fact that we use a different likelihood function
(i.e., the half-Gaussian likelihood) for the upper limit analysis.
Note that the optimized $H_0$ in our best fit analysis is also clearly not fully consistent with
the estimate based on local distance measurements ($H_{0}=73.04\pm1.04$ $\mathrm{km\;s^{-1}\;Mpc^{-1}}$;
\citealt{2022ApJ...934L...7R}) nor the value inferred from {\it Planck}
($H_{0}=67.4\pm0.5$ $\mathrm{km\;s^{-1}\;Mpc^{-1}}$; \citealt{2020A&A...641A...6P}).
This may be an indication that our view of FRBs itself may be incomplete so that we
don't fully understand how to interpret the observations yet.
Nevertheless, if a narrower flat prior $\mathrm{DM_{halo}^{MW}}\in[50,\;80]$
$\mathrm{pc\;cm^{-3}}$ is adopted, we do get an optimized value of $H_0$ ($77.1^{+7.0}_{-7.4}$
$\mathrm{km\;s^{-1}\;Mpc^{-1}}$) in line with our upper limit. It is easy to understand that
the adoption of a larger value of $\mathrm{DM_{halo}^{MW}}$ leads to a smaller value of $\mathrm{DM_{IGM}}$,
thus yielding a lower $H_0$ measurement.

\section{Summary and discussion}\label{sec:Summary}
As more and more FRBs have their $\mathrm{DM_{IGM}}$ associated
with well-measured redshifts, they are emerging as potentially useful cosmic probes.
In this work, we have affirmed the viability of using FRBs to constrain the low-redshift
cosmic expansion and shed light on the growing Hubble tension.

We have used the dispersion measures of 24 localized FRBs within the redshift range
$0.0337\leq z\leq 1.016$ to compare the predictions of $\Lambda$CDM and $R_{\rm h}=ct$.
The free parameters of each model have been optimized via a maximization of the likelihood
function. Given the wide flat prior $\mathrm{DM_{halo}^{MW}}\in[5,\;80]$ $\mathrm{pc\;cm^{-3}}$,
we have shown that flat $\Lambda$CDM fits the observations with $-2\ln \mathcal{L}
=308.97$, a Hubble constant $H_0=95.8^{+7.8}_{-9.2}$ $\mathrm{km\;s^{-1}\;Mpc^{-1}}$ and
$\Omega_{\rm m}\sim0.66$ (poorly constrained; only a median value can be estimated
because the prior is uniform over the interval $[0,\;1]$). One point worth emphasizing
is that the $H_0$ measurement is affected dramatically by the $\mathrm{DM_{halo}^{MW}}$
prior. For the same $\mathrm{DM_{halo}^{MW}}$ prior, $R_{\rm h}=ct$ fits these same data
with $-2\ln \mathcal{L}=310.18$, and a Hubble constant $H_0=94.2^{+5.6}_{-6.2}$
$\mathrm{km\;s^{-1}\;Mpc^{-1}}$. Both models fit the data quite well. Flat $\Lambda$CDM
has one additional free parameter compared to $R_{\rm h}=ct$, however, so the latter is
favored by the BIC with a likelihood of $\sim73\%$ versus $\sim27\%$.

This result is suggestive, though the current sample of FRBs is clearly
too small for either model to be ruled out yet.
We have therefore constructed two simulated samples in the redshift range $0<z<3$,
one based on a $\Lambda$CDM background cosmology and the other built using $R_{\rm h}=ct$.
Through the analysis of these simulated FRBs, we have estimated that a sample of at least 1,150
localized FRBs would be required to rule out $R_{\rm h}=ct$ at a $3\sigma$ confidence level
if the background cosmology is actually $\Lambda$CDM, while $\sim550$ FRBs would
be necessary to similarly rule out $\Lambda$CDM if the real cosmology is instead $R_{\rm h}=ct$.
The difference in required sample size is caused by $\Lambda$CDM's larger number of free parameters
and, therefore, a severe penalty from the BIC.

We have also adopted a straightforward method of deriving an upper limit to $H_{0}$,
eliminating possible biases due to the presumption of an unknown probability distribution of
the DM contributed by the FRB host galaxy ($\mathrm{DM_{host}}$). Owing to the theoretical IGM portion of
DM ($\mathrm{DM_{IGM}}$) being proportional to $H_0$ at any redshift, restricting
$\mathrm{DM_{ext}}$ to be larger than the IGM portion it contains at any $z$
delimits $H_0$ on the upside. Assuming flat $\Lambda$CDM, with a wide flat prior
on $\Omega_{\rm m}$, we have obtained $H_0<89.0$ $\mathrm{km\;s^{-1}\;Mpc^{-1}}$ at
the 95\% confidence level. The inferred upper limit to $H_0$ is very sensitive to
$\Omega_{\rm m}$, however. The Hubble tension between the high-$z$ and low-$z$
measurements of $H_0$ may be due in part to the different emphasis placed on
$\Omega_{\rm m}$ by the CMB analysis versus its impact locally.

In the $R_{\rm h}=ct$ cosmology, we found an upper limit of $H_0<90.0$
$\mathrm{km\;s^{-1}\;Mpc^{-1}}$, also consistent with the low-$z$ measurement. Our
overall conclusion is that FRBs may become a powerful diagnostic to test the Hubble
tension as their sample size grows with the accurate localization of future events.

Previously, within the context of $\Lambda$CDM, \cite{Hagstotz2022} (hereafter H22),
\cite{2022MNRAS.516.4862J} (hereafter J22), and \cite{Wu2022} (hereafter W22) used
observations of FRBs to measure $H_0$, finding $H_0=62.3\pm9.1$ $\mathrm{km\;s^{-1}\;Mpc^{-1}}$,
$H_0=73^{+12}_{-8}$ $\mathrm{km\;s^{-1}\;Mpc^{-1}}$, and $H_0=68.8^{+5.0}_{-4.3}$
$\mathrm{km\;s^{-1}\;Mpc^{-1}}$, respectively. These estimates are not fully consistent with our result
of $H_0=95.8^{+7.8}_{-9.2}$ $\mathrm{km\;s^{-1}\;Mpc^{-1}}$ at the $1\sigma$ confidence level.
It is thus useful to compare differences in the analysis methods. For the analysis
sample, H22, J22, and W22, respectively, used 9, 16 (including 60 unlocalized FRBs), and 18
localized FRBs to constrain $H_0$, while our analysis focused on a larger sample of 24 localized FRBs.

For $\mathrm{DM_{halo}^{MW}}$, J22 set a prior of 50 $\mathrm{pc\;cm^{-3}}$,
and W22 marginalized $\mathrm{DM_{halo}^{MW}}$ using a Gaussian prior of
$\mathrm{DM_{halo}^{MW}}=65\pm15$ $\mathrm{pc\;cm^{-3}}$ over the range of
$[50,\;80]$ $\mathrm{pc\;cm^{-3}}$. Our work instead adopted a wide flat prior
$\mathrm{DM_{halo}^{MW}}\in[5,\;80]$ $\mathrm{pc\;cm^{-3}}$. Since their assumed
$\mathrm{DM_{halo}^{MW}}$ values are higher, their assumed $\mathrm{DM_{IGM}}$ will be decreased,
which we assess to be the main reason for the lower $H_0$ values derived by J22 and W22.
Note that if we used a narrower flat prior $\mathrm{DM_{halo}^{MW}}\in[50,\;80]$ $\mathrm{pc\;cm^{-3}}$,
our result of $H_0=77.1^{+7.0}_{-7.4}$ $\mathrm{km\;s^{-1}\;Mpc^{-1}}$ would be in good
agreement with previous estimates at the $1\sigma$ confidence level.

For the probability distributions of $\mathrm{DM_{host}}$ and $\mathrm{DM_{IGM}}$, J22, W22,
and our work used an asymmetric penalty to model them. H22 instead used a symmetric penalty for
DM lying away from the mean. They neither considered the asymmetric tail to large-DM values,
nor the relatively sharp lower limit of DM at a given redshift expected for $\mathrm{DM_{ext}}$.
The inferred $H_0$ is very sensitive to the assumed value of the mean $\mathrm{DM_{host}}$ (or
$\ln \mathrm{DM_{host}}$). Moreover, numerical simulations show that the mean $\mathrm{DM_{host}}$
moderately increases with redshift. Neither H22 nor J22, however, considered the redshift
evolution of $\mathrm{DM_{host}}$. H22 assumed $\langle\mathrm{DM_{host}}\rangle=100$ $\mathrm{pc\;cm^{-3}}$
in a normal distribution, while J22 used a lognormal function for $\mu_{\mathrm{host}}$
($\equiv\langle\ln \mathrm{DM_{host}}\rangle$), in which $\mu_{\mathrm{host}}$ is treated as
a constant free parameter. Following W22, we also used the same lognormal function to model
$\mu_{\mathrm{host}}$ based on the simulations of \cite{Zhang2020} with $\mu_{\mathrm{host}}$
evolving with redshift, and varying according to the type of FRB host galaxy.
Note that since the assumed value of $\langle\mathrm{DM_{host}}\rangle$ in H22 is larger than
our adopted $e^{\mu_{\mathrm{host}}}$ values from the simulations of \cite{Zhang2020} and
W22 did not include those FRBs with extreme values of $\mathrm{DM_{obs}}$ (such as FRBs~20190520B
and 20210117A), their assumed $\mathrm{DM_{IGM}}$ will be decreased, which we view as the other
reasons for the lower $H_0$ values derived by H22 and W22.

\begin{acknowledgments}
We would like to thank the anonymous referee for constructive comments that helped
improve our work. We are also grateful to Prof. Fa-Yin Wang for sharing the $\mathrm{DM_{IGM}}-z$
relation derived from the IllustrisTNG simulation with us. This work is partially supported by the
National SKA Program of China (2022SKA0130100), the National Natural Science Foundation of China
(grant No. 12041306), the Key Research Program of Frontier Sciences (grant No. ZDBS-LY-7014)
of Chinese Academy of Sciences, International Partnership Program of Chinese Academy of Sciences
for Grand Challenges (114332KYSB20210018), the CAS Project for Young Scientists in Basic Research
(grant No. YSBR-063), the CAS Organizational Scientific Research Platform for National Major
Scientific and Technological Infrastructure: Cosmic Transients with FAST, the Natural Science
Foundation of Jiangsu Province (grant No. BK20221562), and the Young Elite Scientists
Sponsorship Program of Jiangsu Association for Science and Technology.
\end{acknowledgments}

\bibliographystyle{aasjournal}
\bibliography{refs}{}

\begin{thebibliography}{}
\expandafter\ifx\csname natexlab\endcsname\relax\def\natexlab#1{#1}\fi
\providecommand{\url}[1]{\href{#1}{#1}}
\providecommand{\dodoi}[1]{doi:~\href{http://doi.org/#1}{\nolinkurl{#1}}}
\providecommand{\doeprint}[1]{\href{http://ascl.net/#1}{\nolinkurl{http://ascl.net/#1}}}
\providecommand{\doarXiv}[1]{\href{https://arxiv.org/abs/#1}{\nolinkurl{https://arxiv.org/abs/#1}}}

\bibitem[{{Bannister} {et~al.}(2019){Bannister}, {Deller}, {Phillips},
  {Macquart}, {Prochaska}, {Tejos}, {Ryder}, {Sadler}, {Shannon}, {Simha},
  {Day}, {McQuinn}, {North-Hickey}, {Bhandari}, {Arcus}, {Bennert}, {Burchett},
  {Bouwhuis}, {Dodson}, {Ekers}, {Farah}, {Flynn}, {James}, {Kerr}, {Lenc},
  {Mahony}, {O'Meara}, {Os{\l}owski}, {Qiu}, {Treu}, {U}, {Bateman}, {Bock},
  {Bolton}, {Brown}, {Bunton}, {Chippendale}, {Cooray}, {Cornwell}, {Gupta},
  {Hayman}, {Kesteven}, {Koribalski}, {MacLeod}, {McClure-Griffiths},
  {Neuhold}, {Norris}, {Pilawa}, {Qiao}, {Reynolds}, {Roxby}, {Shimwell},
  {Voronkov}, \& {Wilson}}]{2019Sci...365..565B}
{Bannister}, K.~W., {Deller}, A.~T., {Phillips}, C., {et~al.} 2019, Science,
  365, 565, \dodoi{10.1126/science.aaw5903}

\bibitem[{{Becker} {et~al.}(2011){Becker}, {Bolton}, {Haehnelt}, \&
  {Sargent}}]{Becker2011}
{Becker}, G.~D., {Bolton}, J.~S., {Haehnelt}, M.~G., \& {Sargent}, W. L.~W.
  2011, \mnras, 410, 1096, \dodoi{10.1111/j.1365-2966.2010.17507.x}

\bibitem[{{Bhandari} {et~al.}(2020){Bhandari}, {Sadler}, {Prochaska}, {Simha},
  {Ryder}, {Marnoch}, {Bannister}, {Macquart}, {Flynn}, {Shannon}, {Tejos},
  {Corro-Guerra}, {Day}, {Deller}, {Ekers}, {Lopez}, {Mahony}, {Nu{\~n}ez}, \&
  {Phillips}}]{2020ApJ...895L..37B}
{Bhandari}, S., {Sadler}, E.~M., {Prochaska}, J.~X., {et~al.} 2020, \apjl, 895,
  L37, \dodoi{10.3847/2041-8213/ab672e}

\bibitem[{{Bhandari} {et~al.}(2022){Bhandari}, {Heintz}, {Aggarwal}, {Marnoch},
  {Day}, {Sydnor}, {Burke-Spolaor}, {Law}, {Xavier Prochaska}, {Tejos},
  {Bannister}, {Butler}, {Deller}, {Ekers}, {Flynn}, {Fong}, {James}, {Lazio},
  {Luo}, {Mahony}, {Ryder}, {Sadler}, {Shannon}, {Han}, {Lee}, \&
  {Zhang}}]{Bhandari2022}
{Bhandari}, S., {Heintz}, K.~E., {Aggarwal}, K., {et~al.} 2022, \aj, 163, 69,
  \dodoi{10.3847/1538-3881/ac3aec}

\bibitem[{{Bhardwaj} {et~al.}(2021{\natexlab{a}}){Bhardwaj}, {Kirichenko},
  {Michilli}, {Mayya}, {Kaspi}, {Gaensler}, {Rahman}, {Tendulkar}, {Fonseca},
  {Josephy}, {Leung}, {Merryfield}, {Petroff}, {Pleunis}, {Sanghavi}, {Scholz},
  {Shin}, {Smith}, \& {Stairs}}]{2021ApJ...919L..24B}
{Bhardwaj}, M., {Kirichenko}, A.~Y., {Michilli}, D., {et~al.}
  2021{\natexlab{a}}, \apjl, 919, L24, \dodoi{10.3847/2041-8213/ac223b}

\bibitem[{{Bhardwaj} {et~al.}(2021{\natexlab{b}}){Bhardwaj}, {Gaensler},
  {Kaspi}, {Landecker}, {Mckinven}, {Michilli}, {Pleunis}, {Tendulkar},
  {Andersen}, {Boyle}, {Cassanelli}, {Chawla}, {Cook}, {Dobbs}, {Fonseca},
  {Kaczmarek}, {Leung}, {Masui}, {Mnchmeyer}, {Ng}, {Rafiei-Ravandi}, {Scholz},
  {Shin}, {Smith}, {Stairs}, \& {Zwaniga}}]{Bhardwaj2021b}
{Bhardwaj}, M., {Gaensler}, B.~M., {Kaspi}, V.~M., {et~al.} 2021{\natexlab{b}},
  \apjl, 910, L18, \dodoi{10.3847/2041-8213/abeaa6}

\bibitem[{{Chatterjee} {et~al.}(2017){Chatterjee}, {Law}, {Wharton},
  {Burke-Spolaor}, {Hessels}, {Bower}, {Cordes}, {Tendulkar}, {Bassa},
  {Demorest}, {Butler}, {Seymour}, {Scholz}, {Abruzzo}, {Bogdanov}, {Kaspi},
  {Keimpema}, {Lazio}, {Marcote}, {McLaughlin}, {Paragi}, {Ransom}, {Rupen},
  {Spitler}, \& {van Langevelde}}]{Chatterjee2017}
{Chatterjee}, S., {Law}, C.~J., {Wharton}, R.~S., {et~al.} 2017, \nat, 541, 58,
  \dodoi{10.1038/nature20797}

\bibitem[{{CHIME/FRB Collaboration} {et~al.}(2021){CHIME/FRB Collaboration},
  {Amiri}, {Andersen}, {Bandura}, {Berger}, {Bhardwaj}, {Boyce}, {Boyle},
  {Brar}, {Breitman}, {Cassanelli}, {Chawla}, {Chen}, {Cliche}, {Cook},
  {Cubranic}, {Curtin}, {Deng}, {Dobbs}, {Dong}, {Eadie}, {Fandino}, {Fonseca},
  {Gaensler}, {Giri}, {Good}, {Halpern}, {Hill}, {Hinshaw}, {Josephy},
  {Kaczmarek}, {Kader}, {Kania}, {Kaspi}, {Landecker}, {Lang}, {Leung}, {Li},
  {Lin}, {Masui}, {McKinven}, {Mena-Parra}, {Merryfield}, {Meyers}, {Michilli},
  {Milutinovic}, {Mirhosseini}, {M{\"u}nchmeyer}, {Naidu}, {Newburgh}, {Ng},
  {Patel}, {Pen}, {Petroff}, {Pinsonneault-Marotte}, {Pleunis},
  {Rafiei-Ravandi}, {Rahman}, {Ransom}, {Renard}, {Sanghavi}, {Scholz}, {Shaw},
  {Shin}, {Siegel}, {Sikora}, {Singh}, {Smith}, {Stairs}, {Tan}, {Tendulkar},
  {Vanderlinde}, {Wang}, {Wulf}, \& {Zwaniga}}]{2021ApJS..257...59C}
{CHIME/FRB Collaboration}, {Amiri}, M., {Andersen}, B.~C., {et~al.} 2021,
  \apjs, 257, 59, \dodoi{10.3847/1538-4365/ac33ab}

\bibitem[{{Chittidi} {et~al.}(2021){Chittidi}, {Simha}, {Mannings},
  {Prochaska}, {Ryder}, {Rafelski}, {Neeleman}, {Macquart}, {Tejos},
  {Jorgenson}, {Day}, {Marnoch}, {Bhandari}, {Deller}, {Qiu}, {Bannister},
  {Shannon}, \& {Heintz}}]{2021ApJ...922..173C}
{Chittidi}, J.~S., {Simha}, S., {Mannings}, A., {et~al.} 2021, \apj, 922, 173,
  \dodoi{10.3847/1538-4357/ac2818}

\bibitem[{{Cordes} \& {Lazio}(2002)}]{Cordes2002}
{Cordes}, J.~M., \& {Lazio}, T.~J.~W. 2002, arXiv e-prints, astro.
\newblock \doarXiv{astro-ph/0207156}

\bibitem[{{Deng} \& {Zhang}(2014)}]{Deng2014}
{Deng}, W., \& {Zhang}, B. 2014, \apjl, 783, L35,
  \dodoi{10.1088/2041-8205/783/2/L35}

\bibitem[{{Di Valentino} {et~al.}(2021){Di Valentino}, {Mena}, {Pan},
  {Visinelli}, {Yang}, {Melchiorri}, {Mota}, {Riess}, \&
  {Silk}}]{2021CQGra..38o3001D}
{Di Valentino}, E., {Mena}, O., {Pan}, S., {et~al.} 2021, Classical and Quantum
  Gravity, 38, 153001, \dodoi{10.1088/1361-6382/ac086d}

\bibitem[{{Foreman-Mackey} {et~al.}(2013){Foreman-Mackey}, {Hogg}, {Lang}, \&
  {Goodman}}]{Foreman2013}
{Foreman-Mackey}, D., {Hogg}, D.~W., {Lang}, D., \& {Goodman}, J. 2013, \pasp,
  125, 306, \dodoi{10.1086/670067}

\bibitem[{{Fukugita} {et~al.}(1998){Fukugita}, {Hogan}, \&
  {Peebles}}]{1998ApJ...503..518F}
{Fukugita}, M., {Hogan}, C.~J., \& {Peebles}, P.~J.~E. 1998, \apj, 503, 518,
  \dodoi{10.1086/306025}

\bibitem[{{Gao} {et~al.}(2014){Gao}, {Li}, \& {Zhang}}]{Gao2014}
{Gao}, H., {Li}, Z., \& {Zhang}, B. 2014, \apj, 788, 189,
  \dodoi{10.1088/0004-637X/788/2/189}

\bibitem[{{Gao} {et~al.}(2022){Gao}, {Li}, \& {Gao}}]{2022MNRAS.516.1977G}
{Gao}, R., {Li}, Z., \& {Gao}, H. 2022, \mnras, 516, 1977,
  \dodoi{10.1093/mnras/stac2270}

\bibitem[{{Hagstotz} {et~al.}(2022){Hagstotz}, {Reischke}, \&
  {Lilow}}]{Hagstotz2022}
{Hagstotz}, S., {Reischke}, R., \& {Lilow}, R. 2022, \mnras, 511, 662,
  \dodoi{10.1093/mnras/stac077}

\bibitem[{{Hashimoto} {et~al.}(2021){Hashimoto}, {Goto}, {Lu}, {On}, {Santos},
  {Kim}, {Eser}, {Ho}, {Hsiao}, \& {Lin}}]{Hashimoto2021}
{Hashimoto}, T., {Goto}, T., {Lu}, T.-Y., {et~al.} 2021, \mnras, 502, 2346,
  \dodoi{10.1093/mnras/stab186}

\bibitem[{{Heintz} {et~al.}(2020){Heintz}, {Prochaska}, {Simha}, {Platts},
  {Fong}, {Tejos}, {Ryder}, {Aggerwal}, {Bhandari}, {Day}, {Deller},
  {Kilpatrick}, {Law}, {Macquart}, {Mannings}, {Marnoch}, {Sadler}, \&
  {Shannon}}]{2020ApJ...903..152H}
{Heintz}, K.~E., {Prochaska}, J.~X., {Simha}, S., {et~al.} 2020, \apj, 903,
  152, \dodoi{10.3847/1538-4357/abb6fb}

\bibitem[{{Inoue}(2004)}]{2004MNRAS.348..999I}
{Inoue}, S. 2004, \mnras, 348, 999, \dodoi{10.1111/j.1365-2966.2004.07359.x}

\bibitem[{{Ioka}(2003)}]{2003ApJ...598L..79I}
{Ioka}, K. 2003, \apjl, 598, L79, \dodoi{10.1086/380598}

\bibitem[{{James} {et~al.}(2022{\natexlab{a}}){James}, {Prochaska}, {Macquart},
  {North-Hickey}, {Bannister}, \& {Dunning}}]{2022MNRAS.509.4775J}
{James}, C.~W., {Prochaska}, J.~X., {Macquart}, J.~P., {et~al.}
  2022{\natexlab{a}}, \mnras, 509, 4775, \dodoi{10.1093/mnras/stab3051}

\bibitem[{{James} {et~al.}(2022{\natexlab{b}}){James}, {Ghosh}, {Prochaska},
  {Bannister}, {Bhandari}, {Day}, {Deller}, {Glowacki}, {Gordon}, {Heintz},
  {Marnoch}, {Ryder}, {Scott}, {Shannon}, \& {Tejos}}]{2022MNRAS.516.4862J}
{James}, C.~W., {Ghosh}, E.~M., {Prochaska}, J.~X., {et~al.}
  2022{\natexlab{b}}, \mnras, 516, 4862, \dodoi{10.1093/mnras/stac2524}

\bibitem[{{Keating} \& {Pen}(2020)}]{2020MNRAS.496L.106K}
{Keating}, L.~C., \& {Pen}, U.-L. 2020, \mnras, 496, L106,
  \dodoi{10.1093/mnrasl/slaa095}

\bibitem[{{Kirsten} {et~al.}(2022){Kirsten}, {Marcote}, {Nimmo}, {Hessels},
  {Bhardwaj}, {Tendulkar}, {Keimpema}, {Yang}, {Snelders}, {Scholz},
  {Pearlman}, {Law}, {Peters}, {Giroletti}, {Paragi}, {Bassa}, {Hewitt},
  {Bach}, {Bezrukovs}, {Burgay}, {Buttaccio}, {Conway}, {Corongiu}, {Feiler},
  {Forss{\'e}n}, {Gawro{\'n}ski}, {Karuppusamy}, {Kharinov}, {Lindqvist},
  {Maccaferri}, {Melnikov}, {Ould-Boukattine}, {Possenti}, {Surcis}, {Wang},
  {Yuan}, {Aggarwal}, {Anna-Thomas}, {Bower}, {Blaauw}, {Burke-Spolaor},
  {Cassanelli}, {Clarke}, {Fonseca}, {Gaensler}, {Gopinath}, {Kaspi}, {Kassim},
  {Lazio}, {Leung}, {Li}, {Lin}, {Masui}, {Mckinven}, {Michilli}, {Mikhailov},
  {Ng}, {Orbidans}, {Pen}, {Petroff}, {Rahman}, {Ransom}, {Shin}, {Smith},
  {Stairs}, \& {Vlemmings}}]{Kirsten2022}
{Kirsten}, F., {Marcote}, B., {Nimmo}, K., {et~al.} 2022, \nat, 602, 585,
  \dodoi{10.1038/s41586-021-04354-w}

\bibitem[{{Law} {et~al.}(2020){Law}, {Butler}, {Prochaska}, {Zackay},
  {Burke-Spolaor}, {Mannings}, {Tejos}, {Josephy}, {Andersen}, {Chawla},
  {Heintz}, {Aggarwal}, {Bower}, {Demorest}, {Kilpatrick}, {Lazio}, {Linford},
  {Mckinven}, {Tendulkar}, \& {Simha}}]{2020ApJ...899..161L}
{Law}, C.~J., {Butler}, B.~J., {Prochaska}, J.~X., {et~al.} 2020, \apj, 899,
  161, \dodoi{10.3847/1538-4357/aba4ac}

\bibitem[{{Li} {et~al.}(2019){Li}, {Gao}, {Wei}, {Yang}, {Zhang}, \&
  {Zhu}}]{Li2019}
{Li}, Z., {Gao}, H., {Wei}, J.-J., {et~al.} 2019, \apj, 876, 146,
  \dodoi{10.3847/1538-4357/ab18fe}

\bibitem[{{Li} {et~al.}(2020){Li}, {Gao}, {Wei}, {Yang}, {Zhang}, \&
  {Zhu}}]{Li2020}
{Li}, Z., {Gao}, H., {Wei}, J.~J., {et~al.} 2020, \mnras, 496, L28,
  \dodoi{10.1093/mnrasl/slaa070}

\bibitem[{{Li} {et~al.}(2018){Li}, {Gao}, {Ding}, {Wang}, \& {Zhang}}]{Li2018}
{Li}, Z.-X., {Gao}, H., {Ding}, X.-H., {Wang}, G.-J., \& {Zhang}, B. 2018,
  Nature Communications, 9, 3833, \dodoi{10.1038/s41467-018-06303-0}

\bibitem[{{Lorimer} {et~al.}(2007){Lorimer}, {Bailes}, {McLaughlin},
  {Narkevic}, \& {Crawford}}]{Lorimer2007}
{Lorimer}, D.~R., {Bailes}, M., {McLaughlin}, M.~A., {Narkevic}, D.~J., \&
  {Crawford}, F. 2007, Science, 318, 777, \dodoi{10.1126/science.1147532}

\bibitem[{{Macquart} {et~al.}(2020){Macquart}, {Prochaska}, {McQuinn},
  {Bannister}, {Bhandari}, {Day}, {Deller}, {Ekers}, {James}, {Marnoch},
  {Os{\l}owski}, {Phillips}, {Ryder}, {Scott}, {Shannon}, \&
  {Tejos}}]{Macquart2020}
{Macquart}, J.~P., {Prochaska}, J.~X., {McQuinn}, M., {et~al.} 2020, \nat, 581,
  391, \dodoi{10.1038/s41586-020-2300-2}

\bibitem[{{Manchester} {et~al.}(2005){Manchester}, {Hobbs}, {Teoh}, \&
  {Hobbs}}]{2005AJ....129.1993M}
{Manchester}, R.~N., {Hobbs}, G.~B., {Teoh}, A., \& {Hobbs}, M. 2005, \aj, 129,
  1993, \dodoi{10.1086/428488}

\bibitem[{{Marcote} {et~al.}(2020){Marcote}, {Nimmo}, {Hessels}, {Tendulkar},
  {Bassa}, {Paragi}, {Keimpema}, {Bhardwaj}, {Karuppusamy}, {Kaspi}, {Law},
  {Michilli}, {Aggarwal}, {Andersen}, {Archibald}, {Bandura}, {Bower}, {Boyle},
  {Brar}, {Burke-Spolaor}, {Butler}, {Cassanelli}, {Chawla}, {Demorest},
  {Dobbs}, {Fonseca}, {Giri}, {Good}, {Gourdji}, {Josephy}, {Kirichenko},
  {Kirsten}, {Landecker}, {Lang}, {Lazio}, {Li}, {Lin}, {Linford}, {Masui},
  {Mena-Parra}, {Naidu}, {Ng}, {Patel}, {Pen}, {Pleunis}, {Rafiei-Ravandi},
  {Rahman}, {Renard}, {Scholz}, {Siegel}, {Smith}, {Stairs}, {Vanderlinde}, \&
  {Zwaniga}}]{Marcote2020}
{Marcote}, B., {Nimmo}, K., {Hessels}, J.~W.~T., {et~al.} 2020, \nat, 577, 190,
  \dodoi{10.1038/s41586-019-1866-z}

\bibitem[{{McQuinn}(2014)}]{McQuinn2014}
{McQuinn}, M. 2014, \apjl, 780, L33, \dodoi{10.1088/2041-8205/780/2/L33}

\bibitem[{{Meiksin}(2009)}]{Meiksin2009}
{Meiksin}, A.~A. 2009, Reviews of Modern Physics, 81, 1405,
  \dodoi{10.1103/RevModPhys.81.1405}

\bibitem[{{Melia}(2007)}]{2007MNRAS.382.1917M}
{Melia}, F. 2007, \mnras, 382, 1917, \dodoi{10.1111/j.1365-2966.2007.12499.x}

\bibitem[{{Melia}(2015)}]{Melia:2015}
---. 2015, \apss, 356, 393, \dodoi{10.1007/s10509-014-2211-5}

\bibitem[{{Melia}(2020)}]{Melia2020}
---. 2020, {The Cosmic Spacetime} (Oxford: Taylor \& Francis),
  \dodoi{https://doi.org/10.1201/9781003081029}

\bibitem[{{Melia}(2022)}]{Melia2022}
---. 2022, \pasp, 134, 121001, \dodoi{10.1088/1538-3873/aca51f}

\bibitem[{{Melia} \& {Shevchuk}(2012)}]{2012MNRAS.419.2579M}
{Melia}, F., \& {Shevchuk}, A.~S.~H. 2012, \mnras, 419, 2579,
  \dodoi{10.1111/j.1365-2966.2011.19906.x}

\bibitem[{{Mu{\~n}oz} {et~al.}(2016){Mu{\~n}oz}, {Kovetz}, {Dai}, \&
  {Kamionkowski}}]{2016PhRvL.117i1301M}
{Mu{\~n}oz}, J.~B., {Kovetz}, E.~D., {Dai}, L., \& {Kamionkowski}, M. 2016,
  \prl, 117, 091301, \dodoi{10.1103/PhysRevLett.117.091301}

\bibitem[{{Niu} {et~al.}(2022){Niu}, {Aggarwal}, {Li}, {Zhang}, {Chatterjee},
  {Tsai}, {Yu}, {Law}, {Burke-Spolaor}, {Cordes}, {Zhang}, {Ocker}, {Yao},
  {Wan}, {Feng}, {Niino}, {Bochenek}, {Cruces}, {Connor}, {Jiang}, {Dai},
  {Luo}, {Li}, {Miao}, {Niu}, {Anna-Thomas}, {Sydnor}, {Stern}, {Wang}, {Yuan},
  {Yue}, {Zhou}, {Yan}, {Zhu}, \& {Zhang}}]{Niu2022}
{Niu}, C.~H., {Aggarwal}, K., {Li}, D., {et~al.} 2022, \nat, 606, 873,
  \dodoi{10.1038/s41586-022-04755-5}

\bibitem[{{Petroff} {et~al.}(2019){Petroff}, {Hessels}, \&
  {Lorimer}}]{Petroff2019}
{Petroff}, E., {Hessels}, J.~W.~T., \& {Lorimer}, D.~R. 2019, \aapr, 27, 4,
  \dodoi{10.1007/s00159-019-0116-6}

\bibitem[{{Planck Collaboration} {et~al.}(2020){Planck Collaboration},
  {Aghanim}, {Akrami}, {Ashdown}, {Aumont}, {Baccigalupi}, {Ballardini},
  {Banday}, {Barreiro}, {Bartolo}, {Basak}, {Battye}, {Benabed}, {Bernard},
  {Bersanelli}, {Bielewicz}, {Bock}, {Bond}, {Borrill}, {Bouchet}, {Boulanger},
  {Bucher}, {Burigana}, {Butler}, {Calabrese}, {Cardoso}, {Carron},
  {Challinor}, {Chiang}, {Chluba}, {Colombo}, {Combet}, {Contreras}, {Crill},
  {Cuttaia}, {de Bernardis}, {de Zotti}, {Delabrouille}, {Delouis}, {Di
  Valentino}, {Diego}, {Dor{\'e}}, {Douspis}, {Ducout}, {Dupac}, {Dusini},
  {Efstathiou}, {Elsner}, {En{\ss}lin}, {Eriksen}, {Fantaye}, {Farhang},
  {Fergusson}, {Fernandez-Cobos}, {Finelli}, {Forastieri}, {Frailis},
  {Fraisse}, {Franceschi}, {Frolov}, {Galeotta}, {Galli}, {Ganga},
  {G{\'e}nova-Santos}, {Gerbino}, {Ghosh}, {Gonz{\'a}lez-Nuevo}, {G{\'o}rski},
  {Gratton}, {Gruppuso}, {Gudmundsson}, {Hamann}, {Handley}, {Hansen},
  {Herranz}, {Hildebrandt}, {Hivon}, {Huang}, {Jaffe}, {Jones}, {Karakci},
  {Keih{\"a}nen}, {Keskitalo}, {Kiiveri}, {Kim}, {Kisner}, {Knox},
  {Krachmalnicoff}, {Kunz}, {Kurki-Suonio}, {Lagache}, {Lamarre}, {Lasenby},
  {Lattanzi}, {Lawrence}, {Le Jeune}, {Lemos}, {Lesgourgues}, {Levrier},
  {Lewis}, {Liguori}, {Lilje}, {Lilley}, {Lindholm}, {L{\'o}pez-Caniego},
  {Lubin}, {Ma}, {Mac{\'\i}as-P{\'e}rez}, {Maggio}, {Maino}, {Mandolesi},
  {Mangilli}, {Marcos-Caballero}, {Maris}, {Martin}, {Martinelli},
  {Mart{\'\i}nez-Gonz{\'a}lez}, {Matarrese}, {Mauri}, {McEwen}, {Meinhold},
  {Melchiorri}, {Mennella}, {Migliaccio}, {Millea}, {Mitra},
  {Miville-Desch{\^e}nes}, {Molinari}, {Montier}, {Morgante}, {Moss}, {Natoli},
  {N{\o}rgaard-Nielsen}, {Pagano}, {Paoletti}, {Partridge}, {Patanchon},
  {Peiris}, {Perrotta}, {Pettorino}, {Piacentini}, {Polastri}, {Polenta},
  {Puget}, {Rachen}, {Reinecke}, {Remazeilles}, {Renzi}, {Rocha}, {Rosset},
  {Roudier}, {Rubi{\~n}o-Mart{\'\i}n}, {Ruiz-Granados}, {Salvati}, {Sandri},
  {Savelainen}, {Scott}, {Shellard}, {Sirignano}, {Sirri}, {Spencer},
  {Sunyaev}, {Suur-Uski}, {Tauber}, {Tavagnacco}, {Tenti}, {Toffolatti},
  {Tomasi}, {Trombetti}, {Valenziano}, {Valiviita}, {Van Tent}, {Vibert},
  {Vielva}, {Villa}, {Vittorio}, {Wandelt}, {Wehus}, {White}, {White},
  {Zacchei}, \& {Zonca}}]{2020A&A...641A...6P}
{Planck Collaboration}, {Aghanim}, N., {Akrami}, Y., {et~al.} 2020, \aap, 641,
  A6, \dodoi{10.1051/0004-6361/201833910}

\bibitem[{{Platts} {et~al.}(2019){Platts}, {Weltman}, {Walters}, {Tendulkar},
  {Gordin}, \& {Kandhai}}]{Platts2019}
{Platts}, E., {Weltman}, A., {Walters}, A., {et~al.} 2019, \physrep, 821, 1,
  \dodoi{10.1016/j.physrep.2019.06.003}

\bibitem[{{Prochaska} \& {Zheng}(2019)}]{Prochaska2019a}
{Prochaska}, J.~X., \& {Zheng}, Y. 2019, \mnras, 485, 648,
  \dodoi{10.1093/mnras/stz261}

\bibitem[{{Prochaska} {et~al.}(2019){Prochaska}, {Macquart}, {McQuinn},
  {Simha}, {Shannon}, {Day}, {Marnoch}, {Ryder}, {Deller}, {Bannister},
  {Bhandari}, {Bordoloi}, {Bunton}, {Cho}, {Flynn}, {Mahony}, {Phillips},
  {Qiu}, \& {Tejos}}]{2019Sci...366..231P}
{Prochaska}, J.~X., {Macquart}, J.-P., {McQuinn}, M., {et~al.} 2019, Science,
  366, 231, \dodoi{10.1126/science.aay0073}

\bibitem[{{Qiu} {et~al.}(2022){Qiu}, {Zhao}, {Wang}, {Zhang}, \&
  {Zhang}}]{2022JCAP...02..006Q}
{Qiu}, X.-W., {Zhao}, Z.-W., {Wang}, L.-F., {Zhang}, J.-F., \& {Zhang}, X.
  2022, \jcap, 2022, 006, \dodoi{10.1088/1475-7516/2022/02/006}

\bibitem[{{Ravi} {et~al.}(2019){Ravi}, {Catha}, {D'Addario}, {Djorgovski},
  {Hallinan}, {Hobbs}, {Kocz}, {Kulkarni}, {Shi}, {Vedantham}, {Weinreb}, \&
  {Woody}}]{2019Natur.572..352R}
{Ravi}, V., {Catha}, M., {D'Addario}, L., {et~al.} 2019, \nat, 572, 352,
  \dodoi{10.1038/s41586-019-1389-7}

\bibitem[{{Ravi} {et~al.}(2022){Ravi}, {Law}, {Li}, {Aggarwal}, {Bhardwaj},
  {Burke-Spolaor}, {Connor}, {Lazio}, {Simard}, {Somalwar}, \&
  {Tendulkar}}]{2022MNRAS.513..982R}
{Ravi}, V., {Law}, C.~J., {Li}, D., {et~al.} 2022, \mnras, 513, 982,
  \dodoi{10.1093/mnras/stac465}

\bibitem[{{Riess} {et~al.}(2022){Riess}, {Yuan}, {Macri}, {Scolnic}, {Brout},
  {Casertano}, {Jones}, {Murakami}, {Anand}, {Breuval}, {Brink}, {Filippenko},
  {Hoffmann}, {Jha}, {D'arcy Kenworthy}, {Mackenty}, {Stahl}, \&
  {Zheng}}]{2022ApJ...934L...7R}
{Riess}, A.~G., {Yuan}, W., {Macri}, L.~M., {et~al.} 2022, \apjl, 934, L7,
  \dodoi{10.3847/2041-8213/ac5c5b}

\bibitem[{{Ryder} {et~al.}(2022){Ryder}, {Bannister}, {Bhandari}, {Deller},
  {Ekers}, {Glowacki}, {Gordon}, {Gourdji}, {James}, {Kilpatrick}, {Lu},
  {Marnoch}, {Moss}, {Prochaska}, {Qiu}, {Sadler}, {Simha}, {Sammons}, {Scott},
  {Tejos}, \& {Shannon}}]{2022arXiv221004680R}
{Ryder}, S.~D., {Bannister}, K.~W., {Bhandari}, S., {et~al.} 2022, arXiv
  e-prints, arXiv:2210.04680, \dodoi{10.48550/arXiv.2210.04680}

\bibitem[{{Schwarz}(1978)}]{1978AnSta...6..461S}
{Schwarz}, G. 1978, Annals of Statistics, 6, 461

\bibitem[{{Shao} {et~al.}(2011){Shao}, {Dai}, {Fan}, {Zhang}, {Jin}, \&
  {Wei}}]{2011ApJ...738...19S}
{Shao}, L., {Dai}, Z.-G., {Fan}, Y.-Z., {et~al.} 2011, \apj, 738, 19,
  \dodoi{10.1088/0004-637X/738/1/19}

\bibitem[{{Thornton} {et~al.}(2013){Thornton}, {Stappers}, {Bailes},
  {Barsdell}, {Bates}, {Bhat}, {Burgay}, {Burke-Spolaor}, {Champion}, {Coster},
  {D'Amico}, {Jameson}, {Johnston}, {Keith}, {Kramer}, {Levin}, {Milia}, {Ng},
  {Possenti}, \& {van Straten}}]{Thornton2013}
{Thornton}, D., {Stappers}, B., {Bailes}, M., {et~al.} 2013, Science, 341, 53,
  \dodoi{10.1126/science.1236789}

\bibitem[{{Vagnozzi}(2020)}]{2020PhRvD.102b3518V}
{Vagnozzi}, S. 2020, \prd, 102, 023518, \dodoi{10.1103/PhysRevD.102.023518}

\bibitem[{{Vagnozzi} {et~al.}(2022){Vagnozzi}, {Pacucci}, \&
  {Loeb}}]{2022JHEAp..36...27V}
{Vagnozzi}, S., {Pacucci}, F., \& {Loeb}, A. 2022, Journal of High Energy
  Astrophysics, 36, 27, \dodoi{10.1016/j.jheap.2022.07.004}

\bibitem[{{Verde} {et~al.}(2019){Verde}, {Treu}, \&
  {Riess}}]{2019NatAs...3..891V}
{Verde}, L., {Treu}, T., \& {Riess}, A.~G. 2019, Nature Astronomy, 3, 891,
  \dodoi{10.1038/s41550-019-0902-0}

\bibitem[{{Walters} {et~al.}(2019){Walters}, {Ma}, {Sievers}, \&
  {Weltman}}]{Walters2019}
{Walters}, A., {Ma}, Y.-Z., {Sievers}, J., \& {Weltman}, A. 2019, \prd, 100,
  103519, \dodoi{10.1103/PhysRevD.100.103519}

\bibitem[{{Walters} {et~al.}(2018){Walters}, {Weltman}, {Gaensler}, {Ma}, \&
  {Witzemann}}]{Walters2018}
{Walters}, A., {Weltman}, A., {Gaensler}, B.~M., {Ma}, Y.-Z., \& {Witzemann},
  A. 2018, \apj, 856, 65, \dodoi{10.3847/1538-4357/aaaf6b}

\bibitem[{{Wang} \& {Wei}(2023)}]{2023ApJ...944...50W}
{Wang}, B., \& {Wei}, J.-J. 2023, \apj, 944, 50,
  \dodoi{10.3847/1538-4357/acb2c8}

\bibitem[{{Wang} \& {Wang}(2018)}]{2018A&A...614A..50W}
{Wang}, Y.~K., \& {Wang}, F.~Y. 2018, \aap, 614, A50,
  \dodoi{10.1051/0004-6361/201731160}

\bibitem[{{Wei} {et~al.}(2019){Wei}, {Li}, {Gao}, \& {Wu}}]{Wei2019}
{Wei}, J.-J., {Li}, Z., {Gao}, H., \& {Wu}, X.-F. 2019, \jcap, 2019, 039,
  \dodoi{10.1088/1475-7516/2019/09/039}

\bibitem[{{Wei} \& {Melia}(2022)}]{2022ApJ...928..165W}
{Wei}, J.-J., \& {Melia}, F. 2022, \apj, 928, 165,
  \dodoi{10.3847/1538-4357/ac562c}

\bibitem[{{Wei} {et~al.}(2018){Wei}, {Wu}, \& {Gao}}]{2018ApJ...860L...7W}
{Wei}, J.-J., {Wu}, X.-F., \& {Gao}, H. 2018, \apjl, 860, L7,
  \dodoi{10.3847/2041-8213/aac8e2}

\bibitem[{{Wu} {et~al.}(2022){Wu}, {Zhang}, \& {Wang}}]{Wu2022}
{Wu}, Q., {Zhang}, G.-Q., \& {Wang}, F.-Y. 2022, \mnras, 515, L1,
  \dodoi{10.1093/mnrasl/slac022}

\bibitem[{{Xiao} {et~al.}(2021){Xiao}, {Wang}, \& {Dai}}]{Xiao2021}
{Xiao}, D., {Wang}, F., \& {Dai}, Z. 2021, Science China Physics, Mechanics,
  and Astronomy, 64, 249501, \dodoi{10.1007/s11433-020-1661-7}

\bibitem[{{Xu} \& {Han}(2015)}]{2015RAA....15.1629X}
{Xu}, J., \& {Han}, J.~L. 2015, Research in Astronomy and Astrophysics, 15,
  1629, \dodoi{10.1088/1674-4527/15/10/002}

\bibitem[{{Yang} {et~al.}(2022){Yang}, {Wu}, \& {Wang}}]{2022ApJ...940L..29Y}
{Yang}, K.~B., {Wu}, Q., \& {Wang}, F.~Y. 2022, \apjl, 940, L29,
  \dodoi{10.3847/2041-8213/aca145}

\bibitem[{{Yu} \& {Wang}(2017)}]{Yu2017}
{Yu}, H., \& {Wang}, F.~Y. 2017, \aap, 606, A3,
  \dodoi{10.1051/0004-6361/201731607}

\bibitem[{{Zhang}(2022)}]{2022arXiv221203972Z}
{Zhang}, B. 2022, arXiv e-prints, arXiv:2212.03972.
\newblock \doarXiv{2212.03972}

\bibitem[{{Zhang} {et~al.}(2020){Zhang}, {Yu}, {He}, \& {Wang}}]{Zhang2020}
{Zhang}, G.~Q., {Yu}, H., {He}, J.~H., \& {Wang}, F.~Y. 2020, \apj, 900, 170,
  \dodoi{10.3847/1538-4357/abaa4a}

\bibitem[{{Zhang} \& {Li}(2020)}]{2020ApJ...901..130Z}
{Zhang}, L., \& {Li}, Z. 2020, \apj, 901, 130, \dodoi{10.3847/1538-4357/abb091}

\bibitem[{{Zhang} {et~al.}(2021){Zhang}, {Yan}, {Li}, {Zhang}, \&
  {Wang}}]{Zhang2021}
{Zhang}, Z.~J., {Yan}, K., {Li}, C.~M., {Zhang}, G.~Q., \& {Wang}, F.~Y. 2021,
  \apj, 906, 49, \dodoi{10.3847/1538-4357/abceb9}

\bibitem[{{Zhao} {et~al.}(2020){Zhao}, {Li}, {Qi}, {Gao}, {Zhang}, \&
  {Zhang}}]{Zhao2020}
{Zhao}, Z.-W., {Li}, Z.-X., {Qi}, J.-Z., {et~al.} 2020, \apj, 903, 83,
  \dodoi{10.3847/1538-4357/abb8ce}

\bibitem[{{Zheng} {et~al.}(2014){Zheng}, {Ofek}, {Kulkarni}, {Neill}, \&
  {Juric}}]{Zheng2014}
{Zheng}, Z., {Ofek}, E.~O., {Kulkarni}, S.~R., {Neill}, J.~D., \& {Juric}, M.
  2014, \apj, 797, 71, \dodoi{10.1088/0004-637X/797/1/71}

\bibitem[{{Zhou} {et~al.}(2014){Zhou}, {Li}, {Wang}, {Fan}, \&
  {Wei}}]{Zhou2014}
{Zhou}, B., {Li}, X., {Wang}, T., {Fan}, Y.-Z., \& {Wei}, D.-M. 2014, \prd, 89,
  107303, \dodoi{10.1103/PhysRevD.89.107303}

\end{thebibliography}

\end{document}